\documentclass[12pt]{iopart}

\usepackage{times}

\usepackage{graphicx}

\begin{document}
\pagestyle{myheadings}

\def\lesssim{\mathrel{\hbox{\rlap{\hbox{\lower4pt\hbox{$\sim$}}}\hbox{$<$}}}}
\def\gtrsim{\mathrel{\hbox{\rlap{\hbox{\lower4pt\hbox{$\sim$}}}\hbox{$>$}}}}
\newcommand{\ffrac}[2]{\left(\frac{#1}{#2} \right)}

\title{High energy neutrino and tau airshowers in standard and new physics}

\author{D Fargion}

\address{Physics Department and INFN, Rome University 1,Italy}

\begin{abstract}

High Energy Neutrino may lead to a New High Energy Astronomy.
Neutrino interaction in matter at PeVs- EeVs  may behave
differently from Standard Model predictions because possible TeV
Gravity scenario. While traditional $km^3$ neutrino underground
detector will be uneasy to disentangle the Gravity TeV scenario
any new Horizontal Tau Air-Shower detectors at PeVs $\tau$ energy,
 beyond Mountain Chains, will be greatly enhanced and it
may probe fruit-fully the New TeV Physics imprint. Moreover
observing upcoming and horizontal $\tau$ air-showers (UPTAUS,
HORTAUS) from high mountains toward Mount Chains or along widest
Earth Crust  crown masses at the Horizons edges, Ultra High
Energy, UHE, $\nu_{\tau}$ and its consequent $\tau$ lepton decay
in flight, will greatly amplify the single UHE $\nu_{\tau}$ track
and it will test, at highest energies, huge volumes comparable to
future underground $km^3$ ones; observing UPTAUS and HORTAUs from
higher balloons or satellites at orbit altitudes, at GZK
energies, the Horizontal Crown effective Masses  may even exceed
$150$ $km^3$. These Highest energies ($10^{19}$ eV) $\nu_{\tau}$
astronomy  are tuned to test the needed abundant $\nu$ fluence in
GZK Z-Showering model: in this scenario ZeV Ultra High Energy
neutrino are hitting on relic light ($0.1-5$ eV masses)
anti-neutrinos, clustered in dark hot halos, creating UHE Z
bosons whose decay in flight may be the hadronic secondaries
observed on Earth atmosphere as  UHECR, isotropically spread
along the cosmic edges and clustered toward few correlated BL Lac
sources.


\end{abstract}



\section{Introduction: The need and the rise of a New UHE $\nu$  astronomy}

While Cosmic Rays astronomy is severely blurred by random
terrestrial, solar, galactic and extragalactic  magnetic lenses,
the highest $\gamma$ ray astronomy (above tens TeVs) became more
or less  blind because photon-photon opacity (electron pair
production) at different energy windows. Indeed the Infrared- TeV
opacity as well as a more severe BBR($2.75 K$)-PeV cut-off are
bounding the TeVs -PeV $\gamma$ ray astronomy in a very nearby
cosmic ( or even galactic) volumes.   Therefore rarest TeVs gamma
signals are at present the most extreme  trace of High Energy
Astronomy. However we observe copious cosmic rays at higher ($\gg
10^{15}$eV) energies almost isotropically spread in the sky.
However  UHECR astronomy must arise, because at largest energies
($\geq 10^{19}$ eV) the magnetic cosmic lenses bending are almost
un-effective and UHECR point to their astronomical sources.
Because of it also a parasite UHE $\nu$ astronomy is expected to
be present as a necessary consequence of UHECR interaction and
cut-off by photopion production  on cosmic 2.75 BBR, the well
known Greisen, Zatsepin, Kuzmin GZK
cut-off \cite{Greisen:1966jv},\cite{Zatsepin:1966jv}.
Moreover in a
different scenario, the so called Z-Burst or Z-Shower model,
UHECR above GZK cut-off are originated by UHE $\nu$ scattering
onto relic light $\nu$ clustered as a dark hot halos; in this
scenario UHE Astronomy is  not just a consequence but itself the
cause of UHECR signals.
 Let us remind, among  the $\gamma$ TeVs discoveries, the signals of power-full Jets blazing
to us from Galactic (Micro-Quasars) or extragalactic edges (BL
Lacs). At PeVs energies astrophysical $\gamma$ cosmic rays should
also be present, but, excluded a very rare and elusive Cyg$X3$
event, they have not being up date observed; only upper bounds
are known at PeV energies.  The missing $\gamma$ PeVs sources, as
we mentioned, are very probably absorbed by their own photon
interactions (electron pairs creation) at the source environment
and/or along the photon propagation into the cosmic Black Body
Radiation (BBR) or into other diffused background radiation.
Unfortunately PeVs charged cosmic rays, easily bend and bounded
in a random walk by Galactic magnetic fields, loose their
original directionality and their astronomical relevance; their
tangled trajectory resident time in the galaxy is much longer
($\geq 10^{3}$ - $10^{5}$) than any linear neutral trajectory, as
gamma rays, making the charged cosmic rays  more probable to be
observed by nearly a comparable lenght ratio. However
astrophysical UHE neutrino signals at $10^{13}$eV-$10^{19}$eV (or
even higher GZK energies) are unaffected by any radiation cosmic
opacity and may easily open a very new exciting window toward
Highest Energy sources. Being weakly interacting the neutrinos
are ideal microscope to deeply observe  in their accelerator
(Jet,SN,GRB, Black  Hole) cores. Other astrophysical $\nu$
sources at lower energies ($10^{8}$ eV - $10^{12}$ eV) should
also be present, at least at EGRET fluence level, but their
signals are very probably  drowned by the dominant diffused
atmospheric $\nu$, secondaries of muon secondaries, produced as
pion decays by the same charged (and smeared) UHE cosmic rays
(while hitting terrestrial atmosphere): the so called diffused
atmospheric neutrinos. Indeed the atmospheric neutrinos signal is
being observed and its modulation has inferred the first
conclusive evidence for a neutrino mass and for a neutrino
flavour mixing. At lowest (MeVs) $\nu$ energy windows, the
abundant and steady solar neutrino flux, (as well as the prompt,
but rare, neutrino burst from a nearby Super-Novae (SN 1987A)),
have been, in last twenty years, successfully explored, giving
support to neutrino mass reality. Let us mention that Stellar
evolution, Supernova explosion but in particular Early Universe
had over-produced and kept in thermal equilibrium neutrinos whose
relic presence here today pollute the cosmic spaces either
smoothly (lightest relativistic $\nu$) or in denser clustered Hot
halos ( eVs relic $\nu$ masses). A minimal tiny (above $0.07$ eV)
$\nu$ mass, beyond a Standard Model, are already making their
cosmic energy density component almost two order of magnitude
larger than the corresponding 2.75 K (Black Body Radiation) BBR
radiation density. Moreover, as recently noted their relic
presence may play a key role acting as a calorimeter of ZeV UHE
neutrinos born at cosmic edges, solving the GZK cut-off puzzle:
The so called
    Z-burst or Z-WW Shower model
\cite{Fargion Salis 1997}, \cite{Fargion Mele Salis 1999}, \cite{Weiler
1999}, \cite{Fargion et all. 2001b}.
    Here we concentrate on the
possibility to detect a component of the associated  UHE neutrino
flux astronomy (above PeVs-EeV up to GZK energies) by UHE
$\nu_{\tau}$ interactions in Mountain chains or in Earth Crust
leading to Horizontal or Upward Tau Air-Showers \cite{Fargion
2000-2002}, \cite{Fargion 2002b}, \cite{Fargion 2002c}.


\section{The UHE $\nu$ $km^3$ detectors:~~ AMANDA,~~ ANTARES,~~ ICECUBE.}

The UHE
$10^{13}$eV - $10^{16}$eV $\nu$ 's , being weakly interacting and
rarer, may  be captured mainly inside huge volumes, bigger than
Super-Kamiokande ones; at present  most popular detectors
consider underground ones (Cubic Kilometer Size like
AMANDA-NESTOR) or (at higher energy $10^{19}$eV - $10^{20}$eV) the
widest Terrestrial atmospheric sheet volumes (Auger Array
Telescope or EUSO atmospheric Detectors).
 Underground $km^3$ detection is based mainly on $\nu_{\mu}$ tracks above hundred TeVs
energies, because of their high penetration in matter, leading to
$\mu$ kilometer size lepton tails \cite{Gandhi et al 1998}. Rarest
atmospheric horizontal shower are also expected by $\nu$
interactions in air (and, as we shall discuss, in the Earth
Crust) with more secondary tails. While $km^3$ detectors are
optimal for PeVs neutrino muons, the Atmospheric Detectors
(AUGER-EUSO like) exhibit a minimal threshold at highest ($\geq
10^{19} eV$) energies. The $km^3$ sensibility is more tuned to
Tens TeV -PeV astronomy while AUGER has wider acceptance above
GZK energies. As we shall discuss $\tau$ Air-Shower detector
exhibit also huge acceptance at both energy windows  being
competitive both at PeV as well as  at EeVs energy range as well
as GZK ones; we shall not discuss here the $Km^{3}$ detector as
the ICECUBE project.



 \section{UHECR galactic EeV neutron and EeV-PeV neutrinos astronomy}

 Incidentally just around such EeV ($10^{18} eV$) energies
an associated Ultra High Energy Neutron Astronomy  might be
already observed in anisotropic clustering of UHECR data because
of the relativistic neutrons boosted lifetime, up to galactic
sizes, comparable to our distance from the Galactic center.
Therefore UHE  neutrons at EeV may be a source candidate of the
observed tiny EeV anisotropy in UHECR data. Indeed a $4\%$
galactic anisotropy and clustering in EeV cosmic rays has been
recently emerged by AGASA\cite{AGASA 1999} along our nearby
galactic spiral arm. These data have been confirmed by a South
(Australia) detector(SUGAR) \cite{Bellido et all 2001}. Therefore
AGASA might have already experienced a first UHECR-Neutron
astronomy (UHENA)at a very relevant energy flux ($\sim 10 eV
cm^{-2}s^{-1}$). This EeV-UHENA signals may and must also be
source of at least a comparable parasite ($10^{17}-10^{16}$ eV)
secondary tails of UHE decaying  neutrino $\bar{\nu_{e}}$  from
the same neutron beta decay in flight. Their flavour oscillations
and mixing in galactic or extragalactic flights (analogous to
atmospheric and solar ones) must guarantee the presence of all
lepton flavours nearly at equal foot: $\bar{\nu_{e}}$,
$\bar{\nu_{\mu}}$ $\bar{\nu_{\tau}}$ \cite{Fargion 2000-2002}.
 The latter UHE $\bar{\nu_{\tau}}$ imprint (added to other local astrophysical UHE $\nu$
production) could be already recorded \cite{Fargion 2000-2002} as
Upward and Horizontal Tau Air-Showers  Terrestrial Gamma Flash
(considered as secondaries $\gamma$ of Upward Tau Air-Showers and
Horizontal Tau Air-Showers): UPTAUS and HORTAUS.



 \section{The GZK puzzle and $\nu$ astronomy by Z-shower
onto relic neutrino halo}

 At highest energy edges ($\geq
10^{19}-10^{20} eV$), a somehow correlated New UHE Astronomy is
also expected for charged Cosmic Rays; indeed these UHECR have
such a large rigidity to avoid any bending by random galactic or
extragalactic magnetic fields; being nearly undeflected UHECR
should point toward the original sources showing in sky a new
astronomical map. Moreover such UHECR astronomy is bounded by the
ubiquitous cosmic $2.75 K^{o}$ BBR screening (the well known
Greisen, Zat'sepin, Kuzmin GZK cut-off) limiting its origination
inside a very local ($\leq 20 Mpc$) cosmic volume. Surprisingly,
these UHECR above GZK (already up to day above $60$ events) are
not pointing toward any known nearby candidate source. Moreover
their nearly isotropic arrival distributions underlines and
testify a very possible cosmic origination, in disagreement with
any local (Galactic plane or Halo, Local Group) expected footprint
by GZK cut-off. A very weak Super-Galactic imprint seem to be
present but at low level and already above GZK volume.  This
opened a very hot debate in modern astrophysics known as the GZK
paradox. Possible solutions has been found recently beyond
Standard Model assuming a non-vanishing neutrino mass. Indeed at
such Ultra-High energies, neutrino at ZeV energies ($\geq
10^{21}eVs$) hitting onto  relic cosmological light ($0.1-4 eV$
masses) neutrinos \cite{Dolgov2002} nearly at rest in Dark Hot
Halos (galactic or in Local Group) has the unique possibility to
produce UHE resonant Z bosons (the so called Z-burst or better
Z-Showering scenario). The different channel cross-sections in
Z-WW-ZZ-Shower by scattering on relic neutrinos are in Fig.1 while
 the boosted Z-Shower secondary chain are in Fig.2,Fig.3:
\cite{Fargion Salis 1997}, \cite{Fargion Mele Salis 1999},
\cite{Yoshida  et all 1998}, \cite{Weiler 1999}; for a more
updated scenario see \cite{Fargion et all. 2001b}, \cite{Fodor
Katz Ringwald 2002}. Indeed UHE neutrinos are un-effected by
magnetic fields and by BBR screening; they may reach us from far
cosmic edges with negligible absorption. The UHE Z-shower in its
ultra-high energy nucleonic secondary component may be just the
observed final UHECR event on Earth. This possibility has been
reinforced by very recent correlations (doublets and triplets
events) between UHECR directions with brightest Blazars sources
at cosmic distances (redshift $\geq 0.1$) quite beyond ($\geq 300
Mpc$) any allowed GZK cut-off \cite{Gorbunov Tinyakov Tkachev
Troitsky}, \cite{AGASA 1999}, \cite{Takeda et all}. Therefore
there might be a role for GZK neutrino fluxes, either at very
high fluence as primary in the Z-Showering scenario or, at least,
as (but at lower intensities) necessary secondaries of all those
UHECR primary absorbed in cosmic BBR radiation fields by GZK cut
of. Naturally other solutions as topological defects or primordial
relics decay may play a role as a source of UHECR, but the
observed clustering \cite{AGASA 1999},
\cite{Tinyakov-Tkachev2001}, \cite{Takeda et all}, seems to favor
compact sources possibly overlapping far BL Lac sources \cite{
Gorbunov Tinyakov Tkachev Troitsky}. The most recent evidence for
self-correlations clustering at $10^{19}$, $2\cdot10^{19}$,
$4\cdot10^{19}$ eVs energies observed by AGASA (Teshima, ICRR26
Hamburg presentation 2001) maybe a first reflection of UHECR
Z-Showering secondaries: $p, \bar{p}, n, \bar{n}$ \cite{Fargion
et all. 2001b}. A very recent solution
    beyond the Standard Model (but within Super-Symmetry) consider
    Ultra High Energy Gluinos as the neutral particle bearing UHE
    signals interacting nearly as an hadron in Terrestrial Atmosphere
\cite{Berezinsky 2002};
    this solution has a narrow window  for
    gluino masses allowable (and serious problems in production
    bounds), but it is an alternative that deserves attention. To
    conclude this brief Z-Shower model survey one finally needs to
    scrutiny the UHE $\nu$ astronomy and to test the GZK solution
    within Z-Showering Models by any independent search on Earth for
    such UHE neutrinos traces above PeVs reaching even EeVs-ZeVs
    extreme energies.

\begin{figure}
\centering
\includegraphics[width=.77\textwidth]{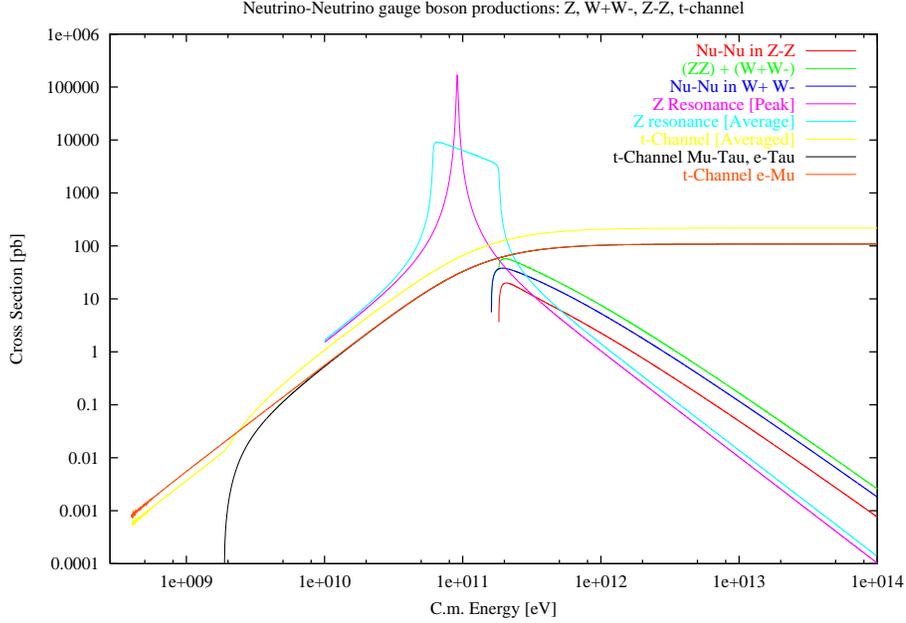}
\caption {The neutrino-relic neutrino cross-sections at center of
mass energy. The Z-peak energy will be smoothed into the
inclined-tower curve, while the WW and ZZ channel will guarantee a
Showering also above a $2 eV$ neutrino masses. The presence of
t-channel play a role in electromagnetic showering at all
energies above the Z-peak.}\label{fig:fig1}
\end{figure}


\begin{figure}
\centering
\includegraphics[width=.77\textwidth]{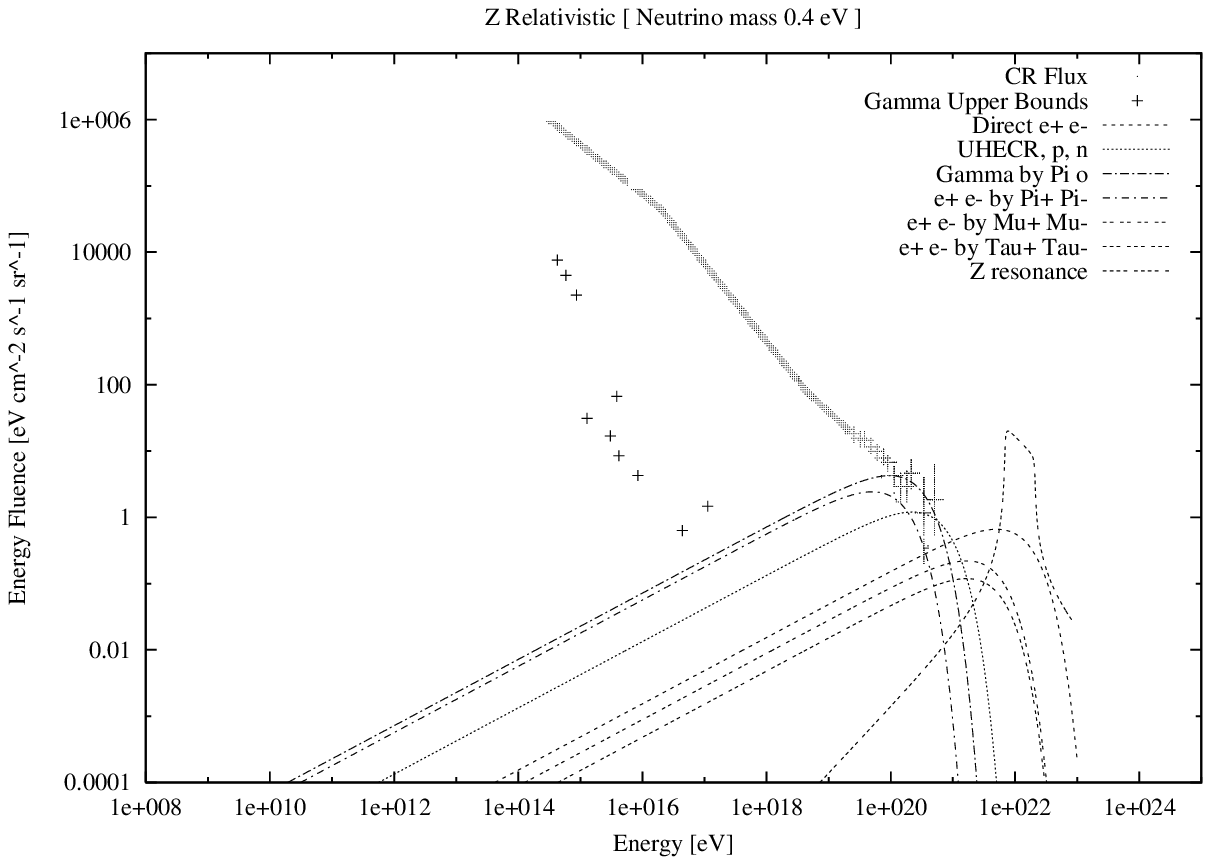}
\caption{Z-Showering Energy Flux distribution for different
    channels assuming a light (fine tuned)  relic neutrino mass
    $m_{\nu} = 0.4 eV$
\cite{Fargion 2000}, \cite{Fargion et all. 2001b},\cite{Fodor
Katz Ringwald 2002}.    Note that the nucleon
    injection energy fit the present AGASA data as well as the very
    recent evidence of a corresponding tiny Majorana neutrino mass
\cite{Klapdor-Kleingrothaus:2002ke}.
    Lighter neutrino masses are able to modulate UHECR at higher energies \cite{Fargion et all. 2001b}.}
\label{fig:fig2a}

\centering
\includegraphics[width=.77\textwidth]{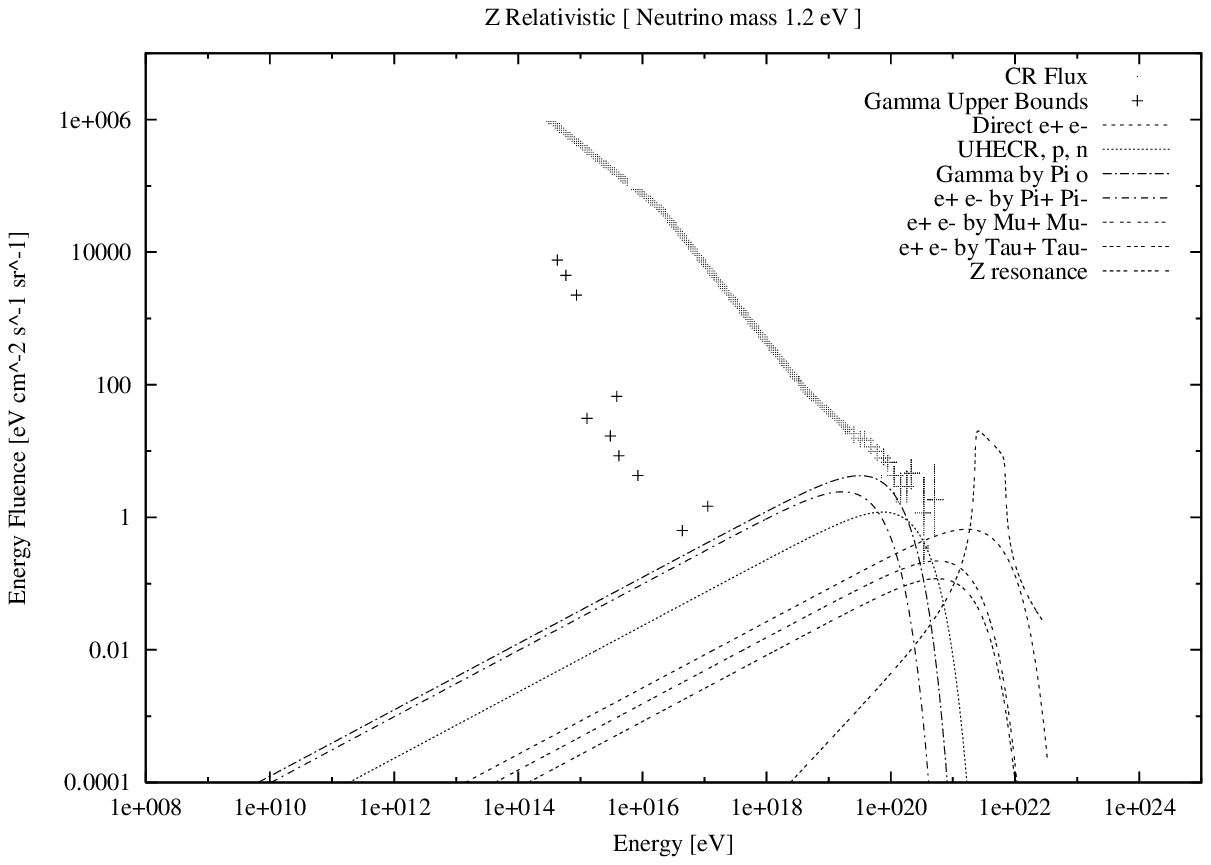}
\caption{Z-Showering Energy Flux distribution for different channels
    assuming a light (fine tuned)  relic neutrino mass $m_{\nu} = 1.2
    eV$ able to partially fill the highest $10^{20} eV$ cosmic ray
    edges. Note that this value lead to a Z-Knee cut-off, above the
    GZK one, well tuned to present Hires data; more details on a very
    light non degenerated neutrino mass ($\leq 0.1 eV$)are discussed
     in the last  figure
\cite{Fargion et all. 2001b}, \cite{Kalashev:2002kx}. }
\label{fig:fig2}
\end{figure}



 \section{ A UHE $\nu$ Astronomy by $\tau$ Air-Shower: An ideal amplifier}

    Recently
\cite{Fargion et all 1999}, \cite{Fargion 2000-2002}
     it has been proposed a new competitive UHE $\nu$ detection
    based on ultra high energy $\nu_{\tau}$
    interaction in matter and its consequent secondary $\tau$ decay
    in flight while escaping from the rock (Mountain Chains, Earth
    Crust) or water (Sea,Ice)  in air leading to Upward or Horizontal
    $\tau$ Air-Showers (UPTAUs and
    HORTAUs),
\cite{Fargion2001a}, \cite{Fargion2001b}.
    In a pictorial
    way one may compare the UPTAUs and HORTAUs as the double bang
    processes expected in $km^3$ ice-water volumes \cite{Learned
    Pakvasa 1995} : the double bang is due first to the UHE
$\nu_{\tau}$ interaction in matter and secondly by its consequent
$\tau$ decay in flight. Here we consider  a (hidden) UHE $\nu$-N
Bang $in$ (the rock-water within a mountain or the Earth Crust)
and a $\tau$ bang $out$ in air, whose shower is better observable
at high altitudes. A similar muon double bang amplifier is not
really occurring because of the extreme decay lenght of
ultarelativistic ($\gtrsim 10^{13}$ eV muons). The main power of
the UPTAUs and HORTAUs detection is the huge amplification of the
UHE neutrino signal, which may deliver almost all its energy in
numerous secondaries traces (Cherenkov lights, gamma, X photons,
electron pairs, collimated muon bundles) in a wider cone volume.
Indeed the multiplicity in $\tau$ Air-showers secondary particles,
$N_{opt} \simeq 10^{12} (E_{\tau} / PeV)$, $ N_{\gamma} (<
E_{\gamma} > \sim  10 \, MeV ) \simeq 10^8 (E_{\tau} / PeV) $ ,
$N_{e^- e^+} \simeq 2 \cdot 10^7 ( E_{\tau}/PeV) $ , $N_{\mu}
\simeq 3 \cdot 10^5 (E_{\tau}/PeV)^{0.85}$ makes easy the
UPTAUs-HORTAUs discover.
 These HORTAUs, also named Skimming neutrinos \cite{Feng et al 2002},
 studied also in peculiar approximation in the frame of AUGER
 experiment,in proximity of Ande Mountain Chains (see Fig. 4)
    \cite{Fargion et all 1999},
\cite{Bertou et all 2002},
    maybe also originated on front of large Vulcano
\cite{Fargion et all 1999}, \cite{Hou Huang 2002}
    either by $\nu_{\tau}N$, $ \bar\nu_{\tau}N$ interactions as
well as by $ \bar\nu_{e} e \rightarrow W^{-} \rightarrow
\bar\nu_{\tau} \tau$. Also UHE $\nu_{\tau}N$, $ \bar\nu_{\tau}N$
at EeVs may be present  in rare AGASA Horizontal Shower (one
single definitive event observed used as un upper bound) facing
Mountain Chain around the the Akeno Array (se Fig.4). This new UHE
$\nu_{\tau}$ detection is mainly based on the oscillated UHE
neutrino $\nu_{\tau}$ originated by more common astrophysical
$\nu_{\mu}$, secondaries of pion-muon decay at PeVs-EeVs-GZK
energies. These oscillations are guaranteed  by Super Kamiokande
evidences for flavour mixing within GeVs atmospheric neutrino data
\cite{Fukuda:1998mi} as well as by most solid and recent
evidences of complete solar neutrino mixing observed by SNO
detector \cite{Bellido et all 2001}.
    Let us remind that HORTAUs (see Fig. 4)
    from mountain chains must nevertheless occur, even
for no flavour mixing, as being inevitable $\bar\nu_{e}$
secondaries of common pion-muon decay chains ($\pi^{-}\rightarrow
\mu^{-}+\bar\nu_{\mu}\rightarrow e^{-}+\bar\nu_{e}$) near the
astrophysical sources at Pevs energies. These Pevs $\bar\nu_{e}$
are mostly absorbed by the Earth and are only rarely arising as
UPTAUS (see Fig. 5 and cross-section in Fig.7). Their Glashow
resonant interaction allow them to be observed as HORTAUs only
within a very narrow and nearby crown edges at horizons (not to be
discussed here).


\begin{figure}
\centering
\includegraphics[width=13cm]{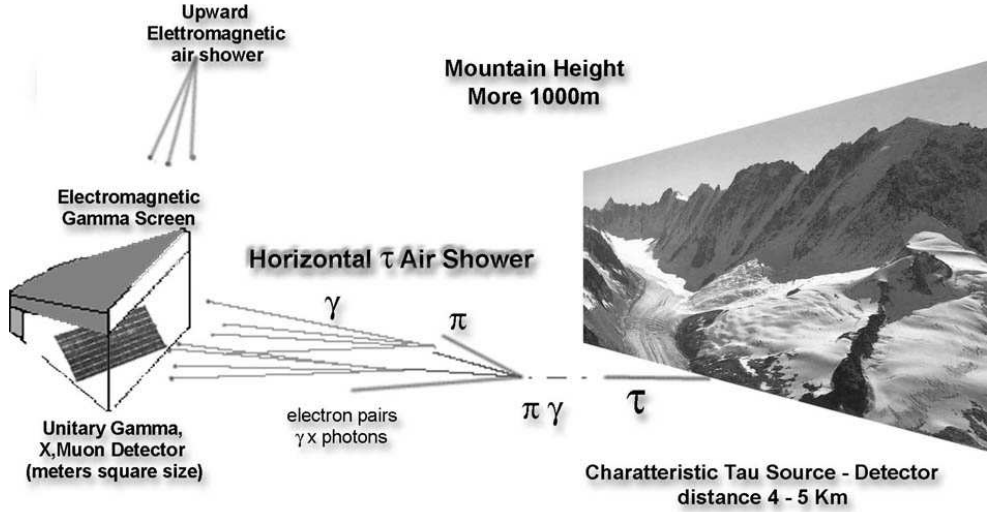}
\vspace{0.5cm}
\caption {The Horizontal Tau on front of a Mountain Chain;
different interaction lenghts will reflect in different event
rate  \cite{Fargion et all 1999}, \cite{Fargion 2000-2002}.}
\label{fig:fig3}
\end{figure}

\begin{figure}
\centering
\includegraphics[width=13cm]{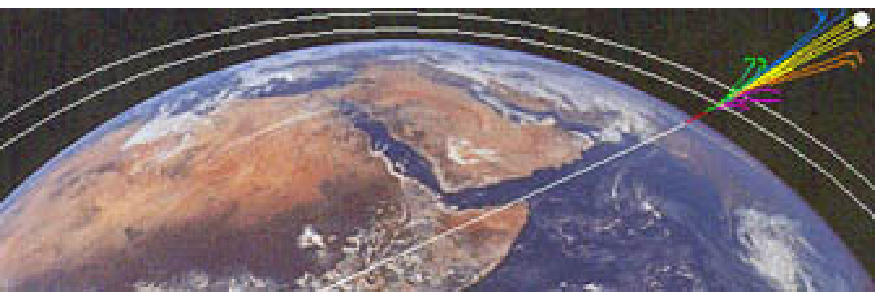}

\vspace{0.5cm}
\caption {The Upward Tau Air-Shower UPTAU and its open fan-like
jets due to geo-magnetic bending at high quota. The gamma Shower
is pointing to an orbital satellite detector as old GRO-BATSE or
very recent Integral one \cite{Fargion 2000-2002},
\cite{Fargion2001a},  \cite{Fargion2001b}.} \label{fig:fig4}
%
%
\vspace{0.5cm}
\centering
\includegraphics[width=13cm]{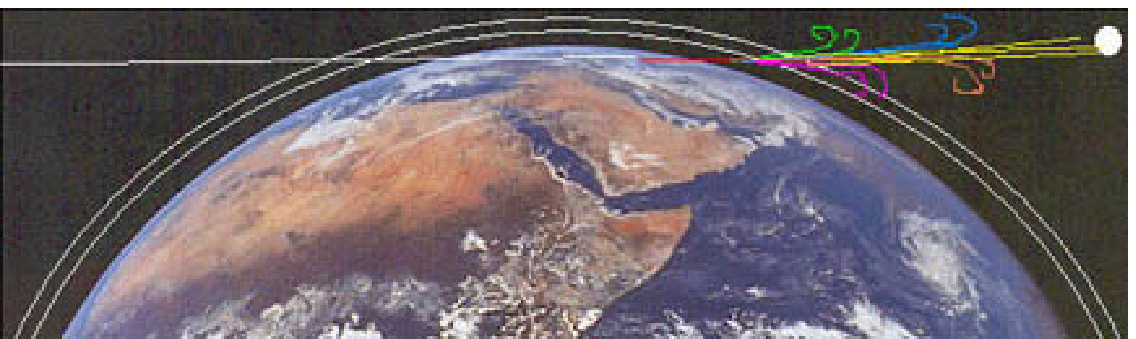}

\vspace{0.5cm}
\caption {The Horizontal-Shower HORTAU and its open fan-like jets
due to geo-magnetic bending at high quota. The gamma Shower is
pointing to an orbital satellite detector as old GRO-BATSE or
very recent Integral one just at the horizons \cite{Fargion
2000-2002}, \cite{Fargion2001a}, \cite{Fargion2001b}.}
\label{fig:fig5}
\end{figure}


At wider energies windows ($10^{14}eV- 10^{20}eV$) only neutrino
$\nu_{\tau}$, $\bar{\nu}_{\tau}$ play a key role in UPTAUS and
HORTAUS. These Showers might be easily detectable looking
downward the Earth's surface from mountains, planes, balloons or
satellites observer. Here the Earth itself acts as a "big
mountain" or a wide beam dump target (see Figs.5-6).
 The present upward $\tau$
at horizons should not be confused with an independent and well
known, complementary (but rarer) Horizontal Tau Air-shower
originated inside the same terrestrial atmosphere: we may referee
to it as the Atmospheric Horizontal Tau Air-Shower. These rare
events are responsible for very rare double bang in air. Their
probability to occur, as derived  in  detail in next paragraph (
summarized in last row Table 1 below, labeled as ratio R between
the event the ground over air event rate) is more than two order
magnitude below the event rate of HORTAUs. The same UPTAUS
(originated in Earth Crust)  have a less competitive upward
showering due to $\nu_{e}$ $\bar\nu_{e}$ interactions within
atmosphere, showering in thin upward air layers \cite{Berezinsky
1990}: this atmospheric Upward Tau presence is as a very small
additional contribute, because rock is more than $3000$ times
denser than air, see Table 1. Therefore at different heights we
need to estimate (See for detail next paragraph) the UPTAUS and
HORTAUs event rate occurring along the thin terrestrial crust
below the observer, keeping care of their correlated  variables:
from a very complex sequence of functions we shall be able to
define and evaluate the effective HORTAUs volumes keeping care of
the thin shower beaming angle, atmosphere opacity and detector
thresholds. At the end of the study, assuming any given neutrino
flux, one might be easily able to estimate at each height $h_{1}$
the expected event rate and the ideal detector size and
sensibility for most detection techniques (Cherenkov,
photo-luminescent, gamma rays, X-ray, muon bundles)(see Fig.18).
The Upcoming Tau Air-Showers and Horizontal ones may be already
recorded as Terrestrial Gamma Flashes as shown by their partial
Galactic signature shown in Fig.8 (over EGRET celestial
background) and in Fig.9 (over EeV anisotropy found by Agasa).

\begin{table}[!b]
\begin{center}
\newcommand{\m}{\hphantom{$-$}}
\renewcommand{\arraystretch}{1.2}
\setlength\tabcolsep{1.2pt}
\caption{\label{Fig18}The Table of the main parameters leading to the
    effective HORTAUs Mass  from the observer height $h_1$, the
    corresponding $\tau$ energy $E_{\tau}$ able to let the $\tau$
    reach him from the horizons, the Total Area $A_{TOT}$ underneath
    the observer, the corresponding $\tau$ propagation lenght in
    matter $l_{\tau}$, the opening angle toward the Crown from the
    Earth $\delta\tilde{\theta}_{h_1}$
    and  $l_{\tau}$ just orthogonal
    in the matter $l_{\tau_{\downarrow}}=l_{\tau}\cdot\sin\delta{\tilde{\theta}_{h_{1}}}$, the Ring Areas for
    two densities $A_R$ at characteristic high altitudes $h_1$, the
    corresponding effective Volume $V_{eff.}$ and the consequent Mass
$\Delta M_{eff.}$ (within the narrow $\tau$ Air-Shower solid
    angle) as a function of density $\rho$ and height$h_1$. In the
    last Row the Ratio R $= M_T/M_{ATM}$ define the ratio of
    HORTAUs produced within the Earth Crown Skin over the atmospheric
    ones: this ratio nearly reflects the matter over air density and
    it reaches nearly two order of magnitude and describe the larger
    probability to observe an HORTAU over the probability to observe
    a double bang in air at EUSO or OWL detectors.}
\vspace{0.5cm}
\begin{tabular}{c||cc|cc|cc|cc}
\hline\hline
    $\rho$  &   1   &   2.65    &   1   &   2.65    &   1   &   2.65    &   1   &   2.65    \\ \hline
    $h_1$   &   2   &   2   &   5   &   5   &   25  &   25  &   500 &   500 \\
    $E_{th}$    &&&&&&&&\\
    (eV)    &   $3.12\cdot 10^{18}$ &   $3.12\cdot 10^{18}$ &   $4.67\cdot 10^{18}$ &   $4.67\cdot 10^{18}$ &   $8\cdot 10^{18}$    &   $8\cdot 10^{18}$    &   $1.08\cdot 10^{18}$ &   $1.08\cdot 10^{18}$ \\
    $A_{TOT}$   &&&&&&&&\\
    $(km^2)$    &   $8\cdot 10^{4}$ &   $8\cdot 10^{4}$ &   $2\cdot 10^{5}$ &   $2\cdot 10^{5}$ &   $10^{6}$    &   $10^{6}$    &   $1.8\cdot 10^{7}$   &   $1.8\cdot 10^{7}$   \\
    $l_\tau$&&&&&&&&\\
    (km)    &   21.7    &   11  &   24.3    &   12.1    &   27.5    &   13.1    &   29.4    &   13.8    \\
    $\delta{\tilde{\theta}}$    &   $1.31^{o}$  &   $0.97^{o}$  &   $1.79^{o}$  &   $1.07^{o}$  &   $2.36^{o}$  &   $1.08^{o}$  &   $2.72^{o}$  &   $1.07^{o}$  \\
    {\footnotesize $l_\tau sin\delta{\tilde{\theta}}$}  &&&&&&&&\\
    (km)    &   0.496   &   0.186   &   0.76    &   0.225   &   1.13    &   0.247   &   1.399   &   0.257   \\
    $A_R$   &&&&&&&&\\
    (Km$^2$)    &   $7.9\cdot 10^{4}$   &   $7.2\cdot 10^{4}$   &   $1.9\cdot 10^{5}$   &   $1.45\cdot 10^{5}$  &   $7.16\cdot 10^{5}$  &   $3.83\cdot 10^{5}$  &   $4.3\cdot 10^{6}$   &   $1.75\cdot 10^{6}$  \\
    d$\Omega/\Omega$    &   $2.5\cdot 10^{-5}$  &   $2.5\cdot 10^{-5}$  &   $2.5\cdot 10^{-5}$  &   $2.5\cdot 10^{-5}$  &   $2.5\cdot 10^{-5}$  &   $2.5\cdot 10^{-5}$  &   $2.5\cdot 10^{-5}$  &   $2.5\cdot 10^{-5}$  \\
    $\Delta$V   &&&&&&&&\\
    (km$^3$)    &   $3.95\cdot 10^{4}$  &   $1.34\cdot 10^{4}$  &   $1.45\cdot 10^{5}$  &   $3.2\cdot 10^{4}$   &   $8.12\cdot 10^{5}$  &   $9.5\cdot 10^{4}$   &   $6\cdot 10^{6}$ &   $4.5\cdot 10^{5}$   \\
    V$_{eff}$=  &&&&&&&&\\
    {\small $\frac{\Delta V\Delta\Omega}{\Omega}$}  &   0.987   &   0.335   &   3.64    &   0.82    &   20.3    &   2.4 &   150.6   &   11.3    \\
    $\Delta$M   &&&&&&&&\\
    (km$^3$)    &   0.987  &    0.89    &   3.64    &   2.17    &   20.3    &   6.3 &   150.6   &   30  \\
    R=  &&&&&&&&\\
    {\footnotesize $\frac{M_T}{M_{ATM}}$}   &   49.6    &   49.2    &   75  &   59.6    &   113 &   65.45   &   140 &   68  \\
&&&&&&&&\\
\hline\hline
\end{tabular}
\end{center}
\end{table}


\begin{figure}
\centering
\includegraphics[width=13cm]{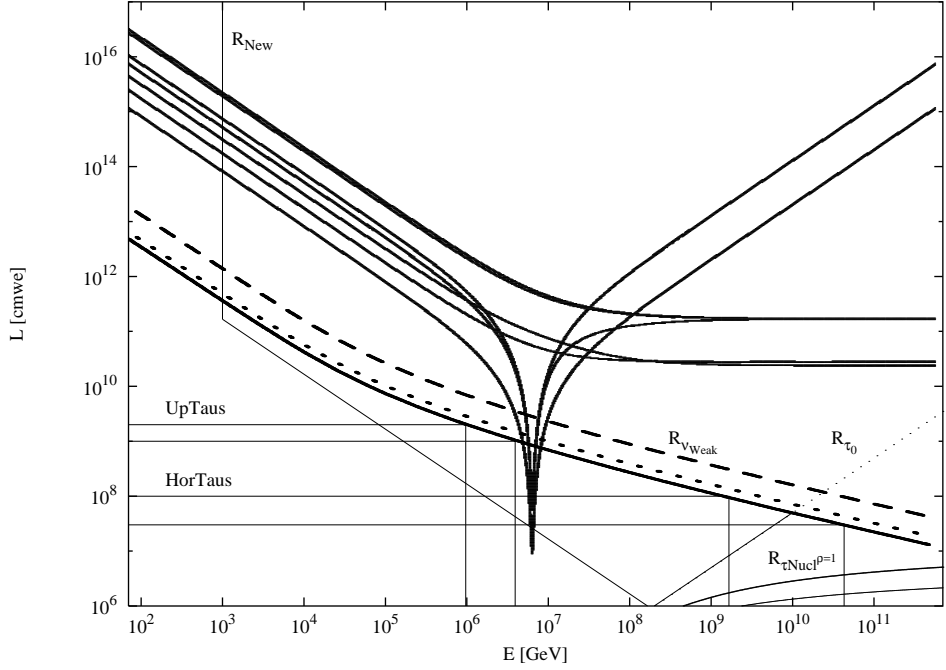}
\caption {Different interaction lenghts for Ultra High Energy
Neutrino that will reflect in different event rate either for
Horizontal Shower from Mountain Chains as well as from Upward and
Horizontal ones from Earth Crust; a severe suppression in  the
neutrino interaction lenght, $R_{New}$, due to any New TeV gravity
will increase by  $2-3$ order of magnitude the neutrino birth
probability leading to an expected tens of thousand event a year
(respect to a hundred a year) by a ten-km lenght Array detector
on front of a Mountain Chain, assuming a $10^3 eV cm^{-2}s^{-1}$
cosmic neutrino fluence (see previous figures) \cite{Fargion
2000-2002}.} \label{fig:fig6}
\end{figure}



\begin{figure}
\centering
\includegraphics[width=13cm]{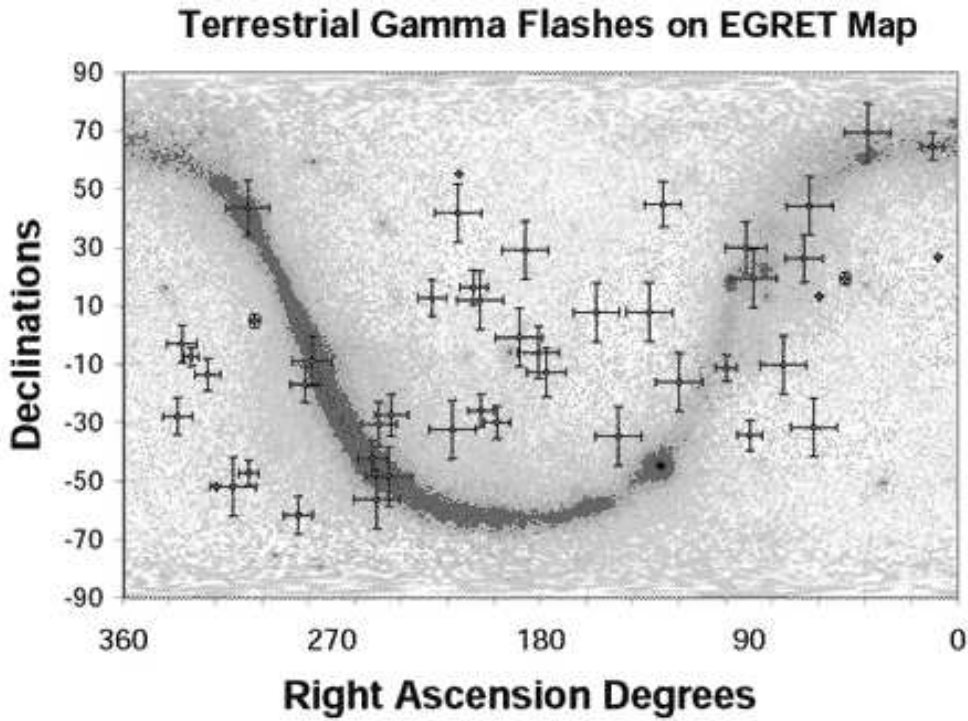}
\caption {The Terrestrial Gamma Flash arrival map over the EGRET
(hundred MeVs-GeV) data in celestial coordinate. It is manifest
the partial galactic signature and the crowding of repeater events
toward the Galactic Center. Also some relevant repeater event are
observed toward anti-galactic direction and to  well known
extragalactic source (see next map) and
    \cite{Fargion 2000-2002}.
}\label{fig:fig7}
\end{figure}



\begin{figure}
\centering
\includegraphics[scale=0.5]{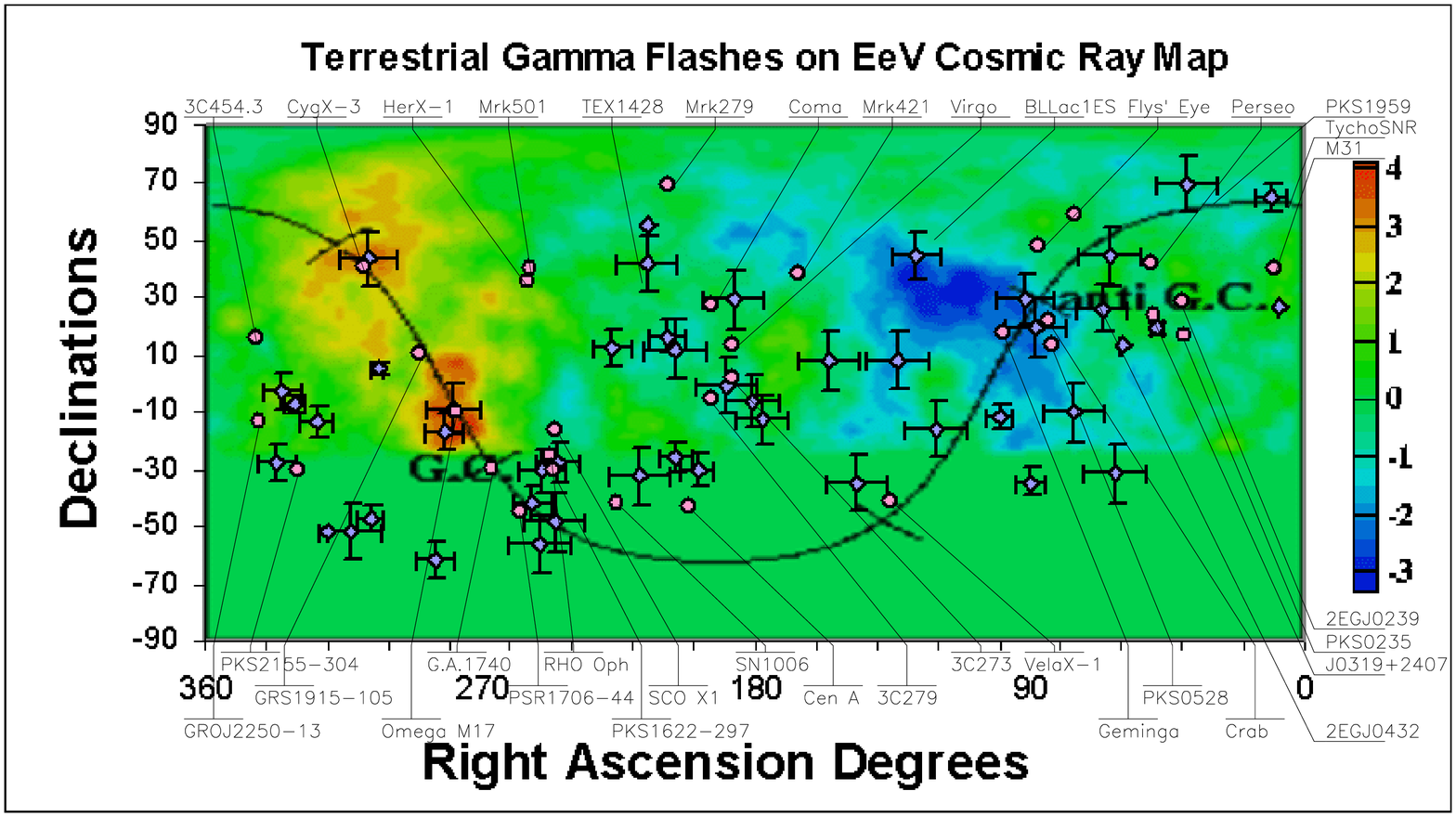}
\caption {The Terrestrial Gamma Flash arrival map over the EeV
anisotropic map  ($ 10^{18} eV$) data in celestial coordinate.
Some relevant X-$\gamma$-TeV sources are also shown in the same
region; see \cite{Fargion 2000-2002}. }\label{fig:fig8}
\end{figure}




\section{The  Ultra High Energy $\nu_{\tau}$ astronomy by UPTAUs-HORTAUs  detection}

The $\tau$ airshowers are observable at different height $h_{1}$
 leading to different underneath observable terrestrial
areas and crust volumes. HORTAUs in deep valley are also relate
to the peculiar geographical morphology and composition
\cite{Fargion 2000-2002} and more in detail as discussed below. We
remind in this case the very important role of UHE  $ \bar\nu_{e}e
\rightarrow W^{-}\rightarrow \bar\nu_{\tau}\tau^{-} $ channels
which may be well observable even in absence of any $\nu_{\tau}$,
$ \bar\nu_{\tau}$ UHE sources or any neutrino flavour mixing: its
Glashow peak resonance make these neutrinos unable to cross all
the Earth across but it may be observable beyond mountain chain
\cite{Fargion 2000-2002}; while testing $\tau$ air-showers beyond
a mountain chain one must consider the possible amplification of
the signal because of a possible New TeV Physics (see
cross-section in Fig 7) \cite{Fargion 2000-2002}. In the following
we shall consider in general the main $\nu_{\tau}-N$, $
\bar\nu_{\tau}-N$ nuclear interaction on Earth crust. It should
be kept in mind also that UPTAUs and in particular HORTAUS are
showering at very low densities and their geometrical escaping
opening angle from Earth (here assumed at far distances
$\theta\sim 1^o$ for rock and $\theta\sim 3^o$ for water) is not
in general conical (like common down-ward showers) but their
ending tails are more shaped in a thin fan-like twin Jets (like
the observed $8$ shaped horizontal Air-Showers) bent and split in
two thin elliptical beams by the geo-magnetic fields. These fan
shape are not widely opened by the Terrestrial Magnetic Fields
while along the North-South magnetic field lines. These
UPTAUs-HORTAUs duration time are also much longer than common
down-ward showers because their showering occurs at much lower
air density and they are more extended: from micro (UPTAUS from
mountains) to millisecond (UPTAUs and HORTAUs from satellites)
long gamma-flashes . Indeed the GRO did observe upcoming
Terrestrial Gamma Flashes are possibly correlated with the UPTAUs
\cite{Fargion 2000-2002}; these events show the expected
millisecond duration times. In order to estimate the rate and the
fluence for of UPTAUs and HORTAUs one has to estimate the
observable crown terrestrial crust mass, facing a complex chain
of questions, leading for each height $h_{1}$, to an effective
observable surface and volume from where UPTAUs and HORTAUs might
be originated. From this effective volume it is easy to estimate
the observable rates, assuming a given incoming UHE $\nu$ flux
model for galactic or extragalactic sources. Here we shall only
refer to the Masses estimate unrelated to any UHE $\nu$ flux
models. These steps are linking simple terrestrial spherical
geometry and its different geological composition, high energy
neutrino physics and UHE $\tau$ interactions, the same UHE $\tau$
decay in flight and its air-showering physics at different quota
within terrestrial air density. Detector physics threshold and
background noises, signal rates have been kept in mind
\cite{Fargion 2000-2002}, but they will be discussed and
explained in  forthcoming papers.


\section{The skin  crown Earth volumes as a function
of $h$ observation height}

Let us therefore define, list and estimate below the sequence of
the key variables whose dependence (shown below or derived in
Appendices) leads to the desired HORTAUs volumes (useful to
estimate the UHE $\nu$ prediction rates) summirized in Table 1
and in Conclusions. These Masses estimate are somehow only a
lower bound that ignore additional contribute by more penetrating
or regenerated $\tau$ \cite{Halzen1998}.  Let us now show the main
functions whose interdependence with the observer altitude lead
to estimate the UPTAUs and HORTAUs equivalent detection Surfaces,
(See Fig. 10-11-13), Volumes and Masses (see Table 1).
\begin{enumerate}
  \item The horizontal distance $d_{h}$ at given height $h_{1}$
  toward the horizons (see Fig. 12):


 \begin{equation}
 d_{h}= \sqrt{( R_{\oplus} + h_1)^2 - (R_{\oplus})^2}= 113\sqrt{ \frac{h_1}{km} }\cdot {km}\sqrt{1+\frac{h_1}{2R_{\oplus}} }
 \end{equation}

The corresponding horizontal edge angle $\theta_{h}$ below the
horizons ($\pi{/2}$) is (see Fig. 11):

\begin{equation}
 {\theta_{h} }={\arccos {\frac {R_{\oplus}}{( R_{\oplus} + h_1)}}}\simeq 1^o \sqrt{\frac {h_{1}}{km}}
\end{equation}

(All the approximations here and below hold for height
$h_{1}\ll{R_{\oplus}}$ )


\begin{figure}
\centering
\includegraphics[width=14cm]{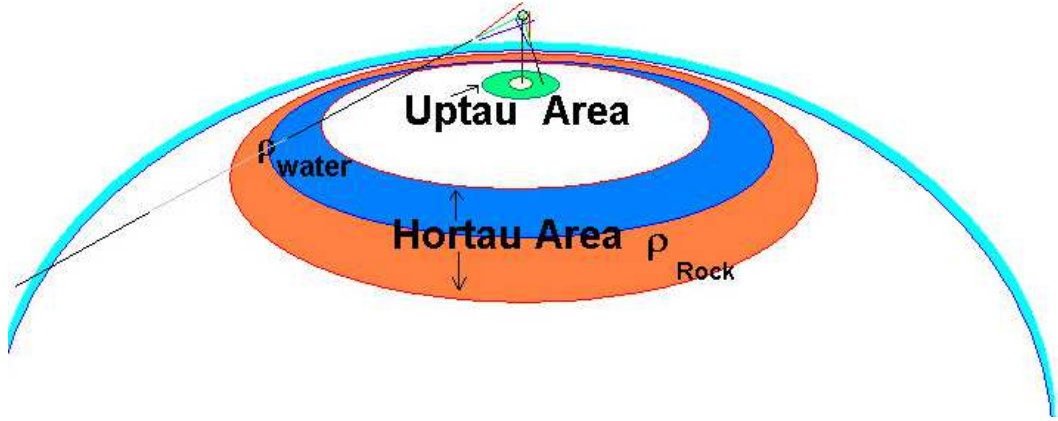}
\caption {The Upward Tau Air-Shower Ring or Crown Areas, either
labeled UPTAUs, the Horizontal Tau Air-Shower Ring Area, labeled
by HORTAU, where the $\tau$ is showering and flashing toward an
observer at height $h_1$. The HORTAU Ring Areas are described both
for water and rock matter density.} \label{fig:fig9}
\end{figure}



 \begin{figure}
\centering
\includegraphics[width=13cm]{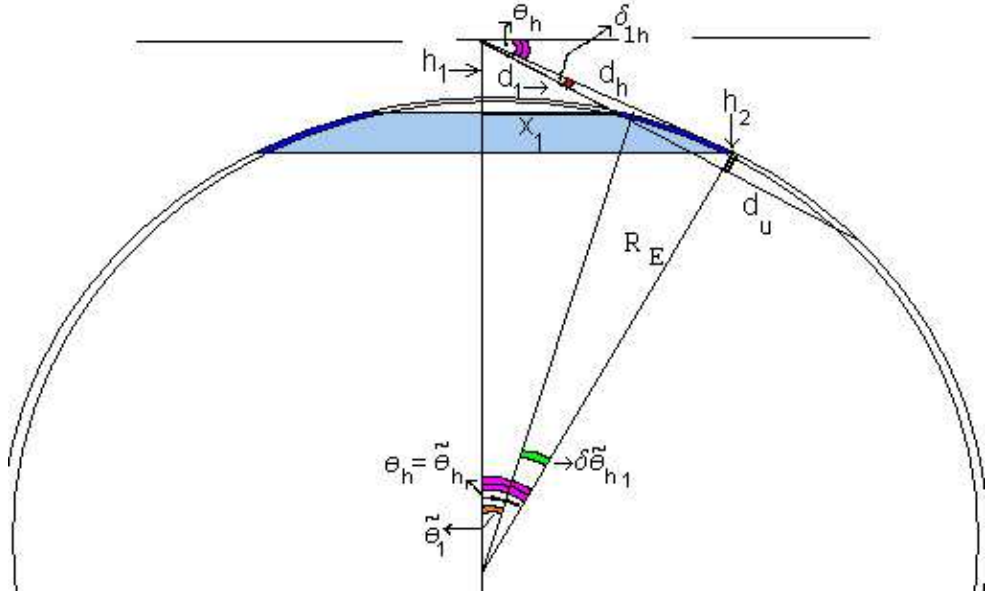}
\caption {The geometrical disposal and the main parameters, as in
the text, defining the UPTAUs and HORTAUs Ring (Crown or Coronas)
Areas; the distances are exaggerated for simplicity.}
\label{fig:fig10}
\end{figure}

  \item
  The consequent characteristic lepton $\tau$ energy
  $E_{\tau_{h}}$ making decay  $\tau$ in flight from  $d_{h}$ distance just nearby
  the source:

\begin{displaymath}
 {E_{\tau_{h}}}=  {\left(\frac {d_{h}}{c\tau_{0}}\right)}m_{\tau} c^2
 \simeq{2.2\cdot10^{18}eV}\sqrt{\frac {h_{1}}{km}}\sqrt{1 + \frac
{h_{1}}{2R_{\oplus}}}
\end{displaymath}


At low quota ($h_1 \leq$ a few kms) the  air depth before the Tau
decay necessary to develop a shower corresponds to a Shower
distance $d_{Sh}$ $\sim 6 kms \ll d_h$.  More precisely at low
quota ($h_1\ll h_o$, where $h_o$ is the air density decay height$
= 8.55$ km.) one finds:
\begin{equation}
d_{Sh}\simeq 5.96km[ 1+ \ln {\frac{E_{\tau}}{10^{18}eV}})] \cdot
e^{\frac{h_1}{h_o}}
\end{equation}
  So we may neglect the distance of the final shower respect to the
longest horizons ones. However at high  altitude ($h_1\geq h_o$)
this is no longer the case (see Appendix A). Therefore  we shall
introduce from here and in next steps a small, but important
modification , whose physical motivation is just to include  the
air dilution role at highest quota: $ {h}_1 \rightarrow \frac
{h_{1}}{1 + h_1/H_o} $, where , as in Appendix A, $H_o= 23$ km.
Therefore previous definition (at height $h_{1} < R_{\oplus}$)
becomes:
\begin{equation}
 {E_{\tau_{h}}}\simeq{2.2\cdot10^{18}eV}\sqrt{\frac {h_{1}}{1 + h_1/H_o}}\sqrt{1 + \frac
{h_{1}}{2R_{\oplus}}}
\end{equation}
This procedure, applied tacitly everywhere, guarantees that there
we may extend our results to those HORTAUs at altitudes where the
residual air density  must exhibit a sufficient slant depth. For
instance, highest $\gg 10^{19} eV$ HORTAUs will be not easily
observable because their ${\tau}$ life distance exceed (usually)
the horizons air depth lenghts.
The parental UHE $\nu_{\tau}$,$\bar\nu_{\tau}$ or $\bar\nu_{e}$
energies $E_{\nu_{\tau}}$ able to produce such UHE $E_{\tau}$ in
matter are:
\begin{equation}
 E_{\nu_{\tau}}\simeq 1.2 {E_{\tau_{h}}}\simeq {2.64
 \cdot 10^{18}eV \cdot \sqrt{\frac {h_{1}}{km}}}
\end{equation}
\item
The neutrino (underground) interaction lenghts  at the
corresponding energies is $L_{\nu_{\tau}}$:
\begin{displaymath}
L_{\nu_{\tau}}= \frac{1}{\sigma_{E\nu_{\tau}}\cdot N_A\cdot\rho_r}
= 2.6\cdot10^{3}km\cdot \rho_r^{-1}{\left(\frac{E_{\nu_h}}{10^8
\cdot GeV}\right)^{-0.363}}
\end{displaymath}
\begin{equation}
{\simeq 304 km \cdot
\left(\frac{\rho_{rock}}{\rho_r}\right)}\cdot{\left(\frac{h_1}{km}\right)^{-0.1815}}
\end{equation}
 For more details see \cite{Gandhi et al 1998}, \cite{Fargion 2000-2002}.
 It should be remind that here we ignore the $\tau$ multi-bangs \cite{Halzen1998}
 that reduce the primary $\nu_{\tau}$ energy and pile up the lower
 energies HORTAUs (EeV-PeVs).

The maximal neutrino depth $h_{2}(h_{1})$, see Fig. 11 above,
under the chord along the UHE neutrino-tau trajectory of lenght
$L_{\nu}(h_{1})$ has been found:
\begin{displaymath}
h_{2}(h_{1}) = {\frac{L_{\nu_{h}}^2}{2^2\cdot{2}(R-h_{2})}
\simeq\frac{L_{\nu_{h}}^2}{8R_{\oplus}}}\simeq 1.81\cdot km
\cdot{\left(\frac{h_1}{km}\right)^{-0.363}\cdot
\left(\frac{\rho_{rock}}{\rho_r}\right)^2}
\end{displaymath}

See Figure 11, for more details. Because the above $h_2$ depths
are in general not  too deep respect to the Ocean depths, we shall
consider respectively either sea (water) or rock (ground)
materials as Crown matter density.
\item
The corresponding opening angle observed from height $h_{1}$,
$\delta_{1h}$ encompassing the underground height  $h_{2}$ at
horizons edge (see Fig. 11) and the nearest UHE $\nu$ arrival
directions $\delta_{1}$:
\begin{displaymath}
{{\delta_{1h}}(h_{2})}={2\arctan{\frac{h_{2}}{2 d_{h}}}}=
2\arctan\left[\frac{{8\cdot
10^{-3}}\cdot{(\frac{h_{1}}{km}})^{-0.863}\left(\frac{\rho_{rock}}{\rho_r}\right)^2}{{\sqrt{1+{\frac{h_{1}}{2R_{\oplus}}}}}}\right]
\end{displaymath}
\begin{equation}
{\simeq 0.91^{o}
\left(\frac{\rho_{rock}}{\rho_r}\right)^2}\cdot{(\frac{h_{1}}{km}})^{-0.863}
\end{equation}
\item
The underground chord $d_{u_{1}}$ (see Fig. $11-13$) where UHE
$\nu_{\tau}$ propagate and the nearest distance $d_{1}$ for
$\tau$ flight (from the observer toward Earth) along the same
$d_{u_{1}}$ direction, within the angle $\delta_{1h}$ defined
above, angle below the horizons (within the upward UHE neutrino
and HORTAUs propagation line) is:
\begin{equation}
d_{u_{1}}=2\cdot{\sqrt{{\sin}^{2}(\theta_{h}+\delta_{1h})(R_{\oplus}+{h_{1}})^{2}-{d_{h}}^2}}
\end{equation}
 Note that by definition  and by construction $ L_{\nu} \equiv d_{u_{1}}$.
The nearest HORTAUs distance corresponding to this horizontal
edges still transparent to UHE $\tau$ is:
\begin{equation}
{d_{1}(h_{1})}=(R_{\oplus}+h_{1})\sin(\theta_{h}+\delta_{1h})-{\frac{1}{2}}d_{u_{1}}
\end{equation}
 Note also that for height $h_{1}\geq km$ : $$ \frac{d_{u_{1}}}{2}\simeq{(R_{\oplus}+{h_{1}})\sqrt{\delta_{1h}\sin{2\theta_{h}}}}\simeq
{158\sqrt{\frac{\delta_{1h}}{1^o}}\sqrt{\frac{h_{1}}{km}}}km.$$



\begin{figure}
\centering
\includegraphics[width=14cm]{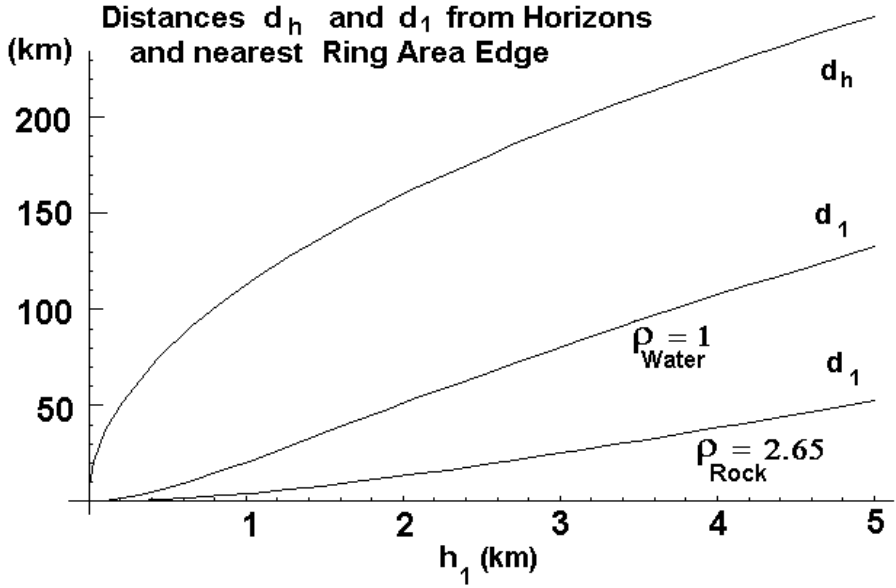}
\caption {Distances from the observer to the Earth ($d_1$) for
    different matter densities or to the Horizons ($d_h$) for low
    altitudes.} \label{fig:fig11}
\end{figure}


\item
The same distance projected cord $x_{1}(h_{1})$ along the
horizontal line (see Fig. 11):
\begin{equation}
x_{1}(h_{1})=d_{1}(h_{1})\cos({\theta_{h}+\delta_{1h}})
\end{equation}

The total terrestrial underneath any observer at height $h_{1}$
is $A_{T}$ (see Figs. 10-13):
$$ A_{T} =2\pi{R_{\oplus}}^{2}(1-\cos{\tilde{\theta}_{h}})
=2\pi{R_{\oplus}}h_{1}\left({\frac{1}{1+\frac{h_{1}}{R_{\oplus}}}}\right)
A_{T}=4\cdot{10}^{4}{km}^{2}{\left({\frac{h_{1}}{km}}\right)}{\left({\frac{1}{1+{\frac{h_{1}}{R_{\oplus}}}}}\right)}$$

Where $\tilde{\theta}_{h}$ is the opening angle from the Earth
along the observer and the horizontal point whose value is the
maximal observable one. At first sight one may be tempted to
consider all the Area  $A_{T}$ for UPTAUs and HORTAUs but because
of the air opacity (HORTAUs) or for its paucity (UPTAUs) this is
incorrect.  While for HORTAUs there is a more complex Area
estimated above and in the following, for UPTAUs the Area Ring (or
Disk) is quite simpler to derive following very similar
geometrical variables summirized in Appendix B.
\item
The Earth Ring Crown crust area ${A_{R}}(h_{1})$ delimited by the
horizons distance $d_{h}$ and the nearest distance $d_{1}$ still
transparent to UHE $\nu_{\tau}$ (see Figs.10-13-16-17). The ring
area ${A_{R}}(h_{1})$ is computed from the internal angles
$\delta{\tilde{\theta}_{h}}$ and $\delta{\tilde{\theta}_{1}}$
defined at the Earth center (Fig.13)(note that
$\delta{\tilde{\theta}_{h}}={\delta{\theta_{h}}}$ but in general
$\delta{\tilde{\theta}_{1}}\neq{\delta{\theta_{1}}}$).

\begin{equation}
\hspace{-2.cm}A_{R}(h_{1})=2\pi{R_{\oplus}}^2(\cos{\tilde{\theta}_{1}}-\cos{\tilde\theta_{h}})
=2\pi{R_{\oplus}^{2}}{\left({{\sqrt{1-{\left({\frac{x_{1}({h_1})}{R_{\oplus}}}\right)^{2}}}}-{\frac{R_{\oplus}}{R_{\oplus}+{h_{1}}}}}\right)}
\end{equation}
 Here $x_{1}({h_1})$ is the cord defined above.



\begin{figure}
\centering
\includegraphics[width=13cm]{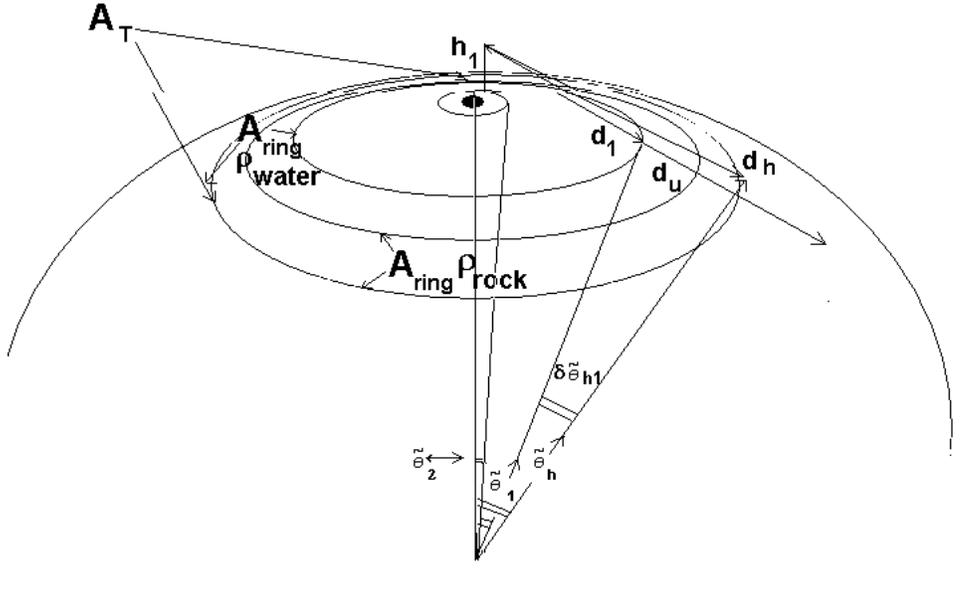}
\caption {Total and Ring (Crown) Areas and Angles for
    UPTAUS-HORTAUS observed at different height.}
\label{fig:fig12}
\end{figure}



\begin{figure}
\centering
\includegraphics[width=15.5cm]{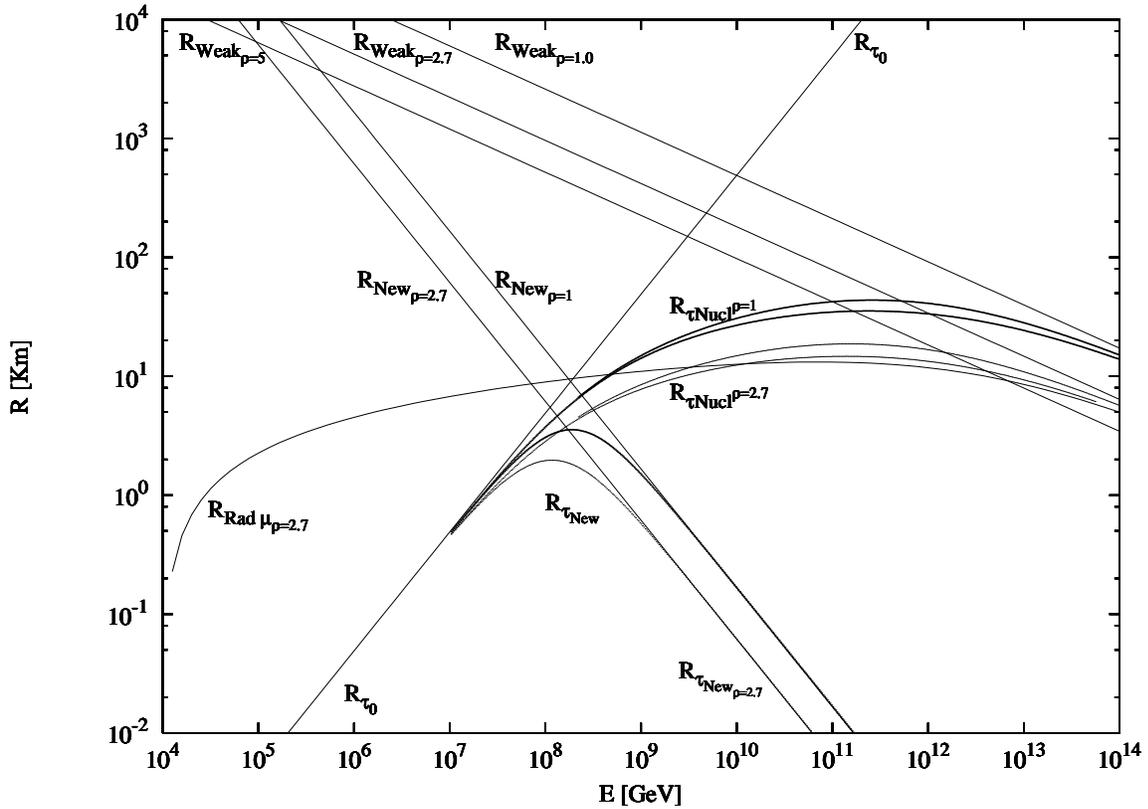}
\caption {Lepton $\tau$ (and $\mu$) Interaction Lenghts for
different matter density: $R_{\tau_{o}}$ is the free $\tau$
lenght,$R_{\tau_{New}}$ is the New Physics TeV Gravity
interaction  range at corresponding
densities,$R_{\tau_{Nucl}\cdot{\rho}}$ ,\cite{Fargion 2000-2002},
see also \cite{Becattini Bottai 2001}, \cite{Dutta et al.2001}, is
the combined $\tau$ Ranges keeping care of all known interactions
and lifetime and mainly the photo-nuclear one. There are two
slightly different split curves (for each density) by two
comparable approximations in the interaction laws.
$R_{Weak{\rho}}$ is the electro-weak Range at corresponding
densities (see also
\cite{Gandhi et al 1998}), \cite{Fargion 2000-2002}.} \label{fig:fig13}
\end{figure}


\item
The characteristic interaction lepton tau lenght $l_{\tau}$
defined at the average $E_{\tau_{1}}$, from interaction in matter
(rock or water). These lenghts have been derived by a analytical
equations keeping care of the Tau lifetime, the photo-nuclear
losses, the electro-weak losses \cite{Fargion 2000-2002}. See
Fig.14 below. The tau lenght along the Earth Skimming distance
$l_{\tau_{2}}$ is vertically projected along the
$\sin(\delta\tilde{\theta}_{h_{1}})$:
\begin{equation}
\delta\tilde{\theta}_{h_{1}}\equiv \tilde\theta_{h}
-\arcsin{\left({\frac{x_{1}}{R_{\oplus}}}\right)}
\end{equation}



\begin{figure}
\centering
\includegraphics[width=13cm]{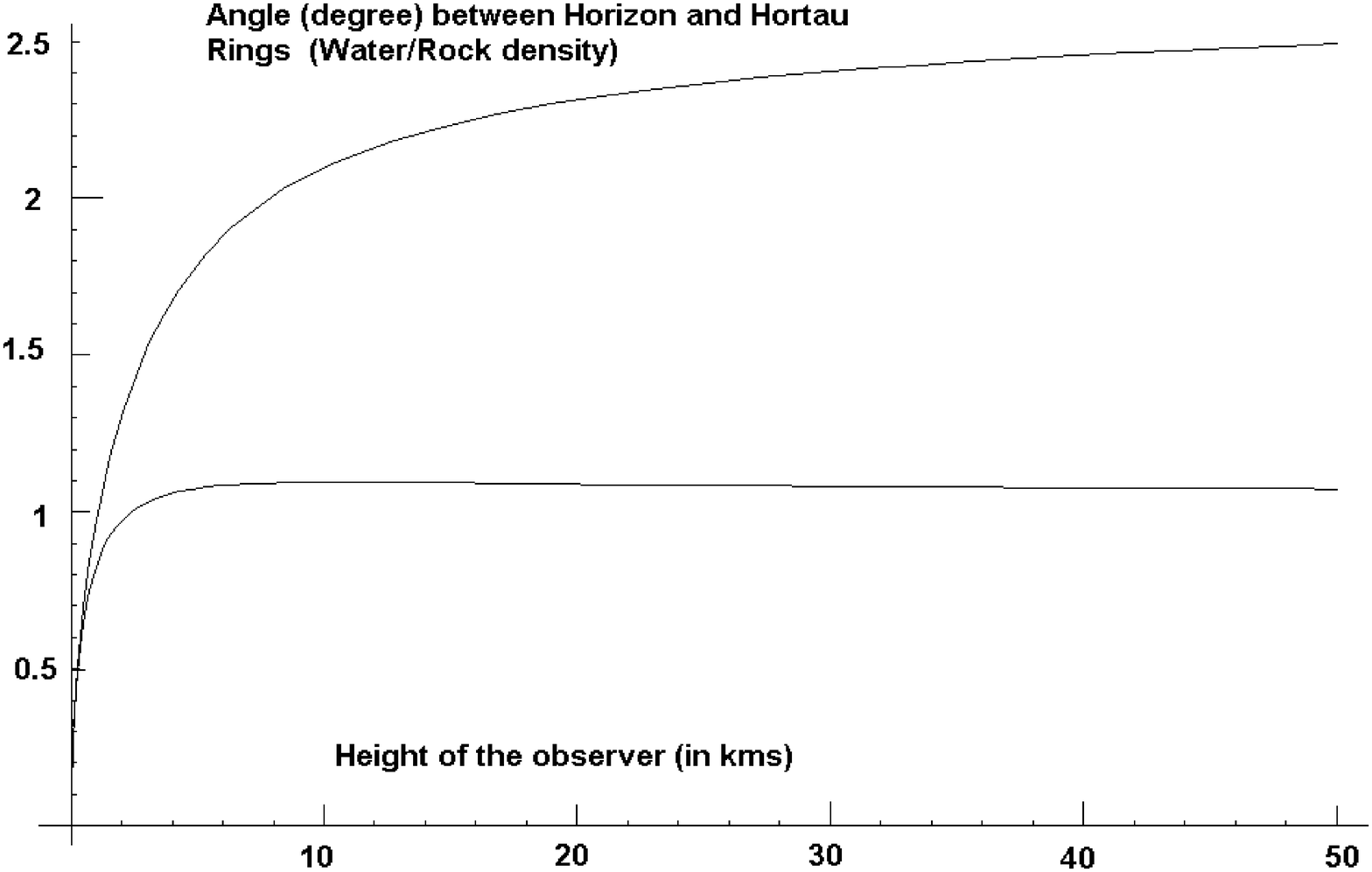}

\vspace{0.5cm} \caption {The $\delta\tilde{\theta}_{h_{1}}$
opening angle toward Ring Earth Skin for density $\rho_{water}$
and $\rho_{rock}$.(see Figs. 11-15) } \label{fig:fig14}
\end{figure}


The same quantity in a more direct approximation:

\begin{displaymath}
\sin\delta{\tilde{\theta}_{h_{1}}}\simeq\frac{L_{\nu}}{2R_{\oplus}}=\frac{{304}km}{2R_{\oplus}}{\left({\frac{\rho_{rock}}{\rho}}\right)}{\frac{h_{1}}{km}}^{-0.1815}.
\end{displaymath}
From highest ($h\gg H_o$=23km) altitude the exact approximation
reduces to:
$$\delta{\tilde{\theta}_{h_{1}}}\simeq{1}^o{\left({\frac{\rho_{rock}}{\rho}}\right)}\left({\frac{h_{1}}{500\cdot
km}}\right)^{-0.1815}$$ Therefore the penetrating $\tau$ skin
depth $l_{\tau_{\downarrow}}$ is
\begin{equation}
l_{\tau_{\downarrow}}=l_{\tau}\cdot\sin\delta{\tilde{\theta}_{h_{1}}}
\simeq{{0.0462\cdot
l_{\tau}{\left({\frac{\rho_{water}}{\rho}}\right)}}}{\frac{h_{1}}{km}}^{-0.1815}
\end{equation}

Where the $\tau$ ranges in matter, $l_{\tau}$ has been calculated
and shown in Fig.14.


\begin{figure}
\centering
\includegraphics[width=16cm]{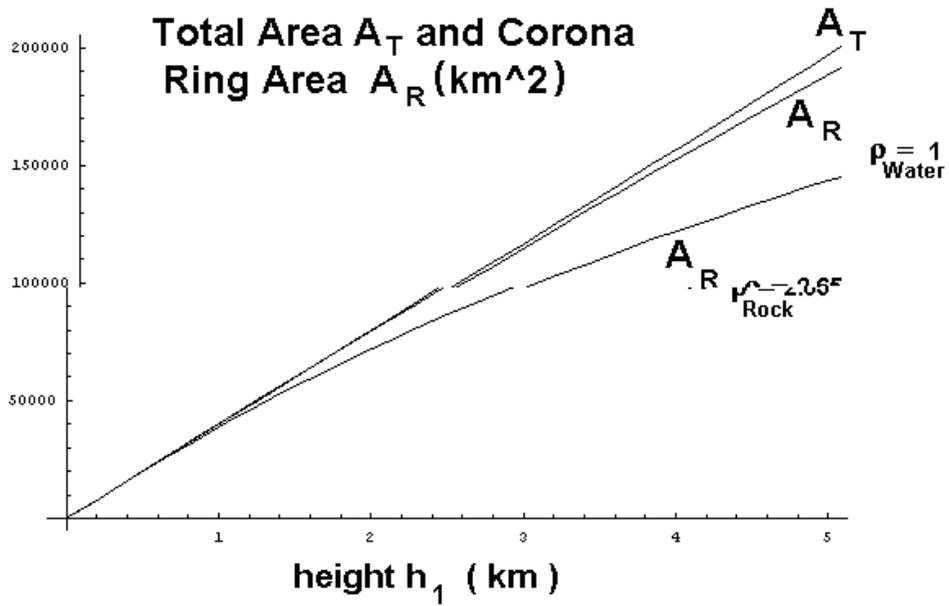}
\caption {Total Area $A_T$ and Ring ( Crown or Coronas) Areas for
two densities $A_R$ at low altitudes. }\label{fig:fig15}
\end{figure}




\begin{figure}
\centering
\includegraphics[width=16cm]{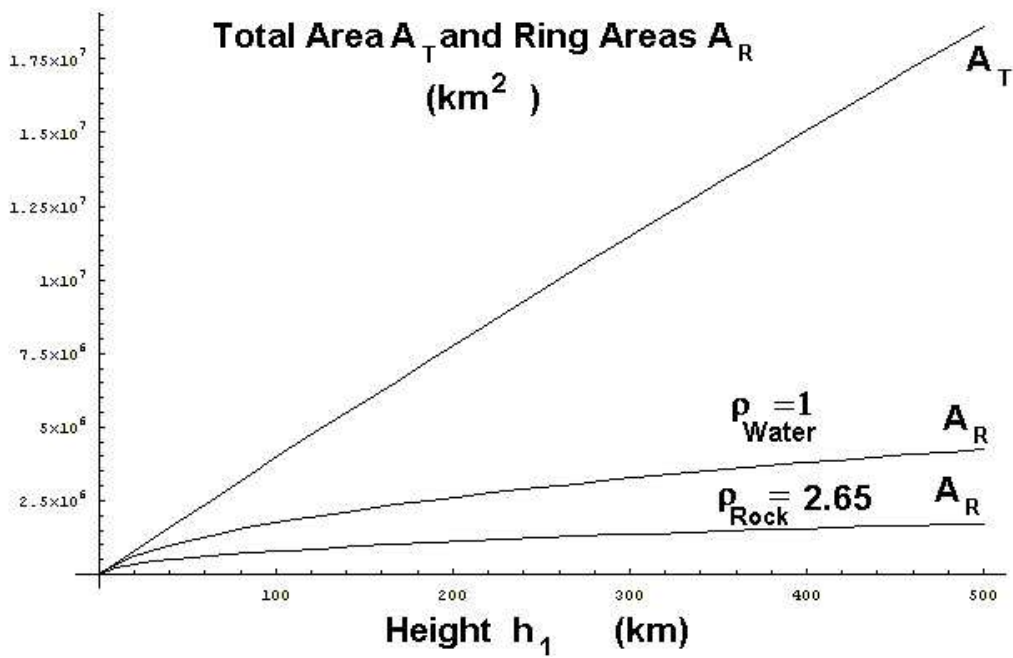}
\caption {Total Area $A_T$ and Ring (Crown) Areas for two
densities $A_R$ at high altitudes. } \label{fig:fi16}
\end{figure}



\item
The final  analytical expression for the Earth Crust Skin Volumes
and Masses under the Earth Skin inspected by HORTAUs are derived
combining the above functions on HORTAUs Areas  with the previous
lepton Tau $l_{\tau_{\downarrow}}$ vertical depth depths:
\begin{equation}
{V_{h_{1}}}={A_{R}}(h_{1})\cdot l_{\tau_{\downarrow}};
\end{equation}
\begin{equation}
{M_{h_{1}}}={V_{h_{1}}}\cdot{\left({\frac{\rho}{\rho_{water}}}\right)}
\end{equation}

\item A More approximated but easy to handle
 expression for Ring area for high altitudes ($h_1\gg 2km$ $h_1\ll R_{\oplus}$) may be
summirized as:
\begin{displaymath}
{A_R (h_1)}\simeq
2\pi{R_{\oplus}^2}\sin{\theta_{h}}{\delta{\tilde{\theta}_{{h}_{1}}}}\propto{\rho^{-1}}
\end{displaymath}
\begin{equation}
\simeq{2\pi{R_{\oplus}^{2}}{\sqrt{\frac{2h_{1}}{R_{\oplus}}}}\left({\frac{\sqrt{1+{\frac{h_{1}}{2R_{\oplus}}}}}{1+{\frac{h_{1}}{R}}}}\right)}{\left({\frac{L_{\nu}}{2R_{\oplus}}}\right)}
\end{equation}
 At high altitudes the above approximation corrected
accordingly to the exact one shown in Figure, becomes:
\begin{equation}
{A_R (h_1)}\simeq
2\pi{R_{\oplus}}{d_{h1}}{\delta{\tilde{\theta}_{{h}_{1}}}}
\simeq{4.65 \cdot 10^6{\sqrt {\frac{h_{1}}{500
km}}}}{\left({\frac{\rho_{water}}{\rho}}\right)}{km}^2
\end{equation}
 Within the above
approximation the final searched Volume ${V_{h_{1}}}$ and Mass
${M_{h_{1}}}$ from where HORTAUs may be generated  is:
\begin{equation}
{V_{h_{1}}}={\frac{\pi}{2}{\sqrt{\frac{2h_{1}}{R_{\oplus}}}}
\left({\frac{\sqrt{1+{\frac{h_{1}}{2R_{\oplus}}}}}{1+{\frac{h_{1}}{R_{\oplus}}}}}\right)}
{L_{\nu}^{2}}{l_{\tau}}\propto{\rho^{-3}}
\end{equation}
\begin{equation}
{M_{h_{1}}}={\frac{\pi}{2}{\sqrt{\frac{2h_{1}}{R_{\oplus}}}}\left({\frac{\sqrt{1+{\frac{h_{1}}{2R_{\oplus}}}}}{1+{\frac{h_{1}}{R_{\oplus}}}}}\right)}{L_{\nu}^{2}}{l_{\tau}}{\rho}\propto{\rho^{-2}}
\end{equation}
\item
The effective observable Skin Tau Mass $M_{eff.}(h_{1})$ within
the thin HORTAU or UPTAUs Shower angle beam $\simeq$ $1^o$ is
suppressed by the solid angle of view
${\frac{\delta\Omega}{\Omega}} \simeq 2.5\cdot 10^{-5}$.
\begin{equation}
{\Delta
M_{eff.}(h_{1})={V_{h_{1}}}\cdot{\left({\frac{\rho}{\rho_{water}}}\right){\frac{\delta\Omega}{\Omega}}}}
\end{equation}
The lower bound Masses $M_{eff.}(h_{1})$ exactly estimated , with
no approximation as in eq.19, from eq.15, at different realistic
high quota experiments, are discussed in the Conclusion below and
summirized in Table 1.


\section{Summary and conclusions}
The discover of the expected UHE neutrino Astronomy is urgent and
just behind the corner. Huge volumes are necessary. Beyond
underground $km^3$ detectors a new generation of UHE neutrino
calorimeter lay on front of mountain chains and just underneath
our feet: The Earth itself  offers huge Crown Volumes as Beam Dump
calorimeters observable via upward Tau Air Showers, UPTAUs and
HORTAUs. Their effective Volumes as a function of the quota $h_1$
has been derived by an analytical function variables in equations
above and Appendix B. These Volumes are discussed below and
summirized in the last column of Table 1 and displayed as bounds
in Fig. 18.
At a few tens meter altitude the UPTAUs and HORTAUs
Ring are almost overlapping. At low altitude $h_{1}\leq 2$ Km the
HORTAUs are nearly independent on the $\rho$ matter density:
${\Delta M_{eff.}(h_{1}=2 Km)(\rho_{Water})}= 0.987 km^3$
${\Delta M_{eff.}(h_{1}=2km)(\rho_{Rock})}= 0.89 km^3$
These volumes are the effective Masses expressed in Water
equivalent volumes. On the contrary at higher quotas, like highest
Mountain observations sites, Airplanes, Balloons and Satellites,
the matter density of the HORTAUs Ring (Crown) Areas play a more
and more dominant role asymptotically $proportional$ to
${\rho}^{-2}$:
${\Delta M_{eff.}(h_{1}=5 km)(\rho_{Water})}= 3.64 km^3 $;
${\Delta M_{eff.}(h_{1}=5 km)(\rho_{Rock})}= 2.17 km^3 $.
 From Air-planes or balloons the effective volumes $M_{eff.}$
increases and the density $\rho$ plays a relevant role.
${\Delta M_{eff.}(h_{1}=25 km)(\rho_{Water})}= 20.3 km^3
$;
${\Delta M_{eff.}(h_{1}=25 km)(\rho_{Rock})}= 6.3 km^3
$. Finally from satellite altitudes the same effective volumes
$M_{eff.}$ are reaching extreme values:
\begin{displaymath}
{\Delta M_{eff.}(h_{1}=500~ km)(\rho_{Water})}= 150.6~ km^3
\end{displaymath}
\begin{displaymath}
{\Delta M_{eff.}(h_{1}=500~ km)(\rho_{Rock})}=30~ km^3
\end{displaymath}
These Masses must be compared with other proposed $km^3$
detectors, keeping in mind that these HORTAUs signals conserve the
original UHE $\nu$ direction information within a degree. One
has  to discriminate  HORTAUS (only while observing from
satellites) from Horizontal High Altitude Showers (HIAS)
\cite{Fargion2001b}, due to rare UHECR showering on high
atmosphere. While wide (RICE)one might also remind the UPTAUs (at
PeVs energies) volumes as derived in Appendix B and in
\cite{Fargion 2000-2002} whose values (assuming an arrival
angle$\simeq 45^o- 60^o$ below the horizons) are nearly
$proportional$ to the $\rho$ density:
\begin{displaymath}
{\Delta M_{eff.}(h_{1}=500~ km)(\rho_{Water})}= 5.9~ km^3
\end{displaymath}
\begin{displaymath}
{\Delta M_{eff.} (h_{1}=500~ km)(\rho_{Rock})}=15.6~ km^3
\end{displaymath}
These widest Masses values, here estimated analytically for main
quota, are offering an optimal opportunity to reveal UHE $\nu$ at
PeVs and EeVs-GZK energies by crown array detectors
(scintillators, Cherenkov, photo-luminescent) facing vertically
the Horizontal edges, located at high mountain peaks or at
air-plane low sides and finally on  balloons and satellites. As
it can be seen in last row of  Table1, the ratio $ R $ between
HORTAUs events and Showers over atmospheric UHE $\nu$ interaction
is a greater and greater number with growing height, implying a
dominant role (above two order of magnitude) of HORTAUS grown in
Earth Skin Crown over Atmospheric HORTAUs. These huge acceptance
may be estimated by comparison with other detector thresholds
(see fig.18 adapted to present GZK-$\nu$ models).


\section{Event rate of upward and horizontal Tau air-showers}

The event rate for HORTAUs are given at first approximation by the
following expression normalized to any given neutrino flux
${\Phi_{\nu}}$:
\begin{equation}
{\dot{N}_{year}}={\Delta
M_{eff.}}\cdot{\Phi_{\nu}}\cdot{\dot{N_o}}\cdot\frac{\sigma_{E_{\nu
}}}{\sigma_{E_{\nu_o}}}
\end{equation}
Where the ${\dot{N_o}}$ is the UHE neutrino rate estimated for
$km^3$ at any given (unitary) energy ${E_{\nu_o} }$, in absence of
any Earth shadow. In our case we shall normalize our estimate at
${E_{\nu_o}=3}$ PeVs energy for standard electro-weak charged
current in a standard parton model \cite{Gandhi et al 1998} and
we shall assume a  model-independent neutrino maximal flux
${\Phi_{\nu}}$ at a flat fluence value of nearly ${\Phi_{\nu}}_o$
$\simeq 3\cdot 10^3 eV cm^{-2}\cdot s^{-1}\cdot sec^{-1}\cdot
sr^{-1}$ corresponding to a characteristic Fermi power law in UHE
$\nu$ primary production rate decreasing as $\frac
{dN_{\nu}}{dE_{\nu}}\simeq {E_{\nu}}^{-2}$ just below present
AMANDA bounds. The consequent rate becomes:
\begin{displaymath}
{\dot{N}_{year}}= 29 {\frac{{\Delta
M_{eff.}}}{km^{3}}\cdot\frac{{\Phi_{\nu}}}{{\Phi_{\nu}}_o}}\cdot\frac{\sigma_{E_{\nu
}}}{\sigma_{E_{\nu_o}}}
\end{displaymath}
\begin{equation}
=29\cdot{\left(\frac{E_{\nu}}{3 \cdot 10^6 \cdot
GeV}\right)^{-0.637}} {\frac{{\Delta
M_{eff.}}}{km^{3}}\cdot\frac{{\Phi_{\nu}}}{{\Phi_{\nu}}_o}}
\end{equation}
For highest satellites and for a characteristic UHE GZK energy
fluence ${\Phi_{\nu}}_o \simeq $
\protect \newline $3 \cdot 10^3 eV cm^{-2}\cdot
s^{-1}\cdot sr^{-1}$(as the needed Z-Showering one), the
consequent event rate observable ${\dot{N}_{year}}$ above the Sea
is :
\begin{equation}
=24.76\cdot{\left(\frac{E_{\nu}}{10^{10} \cdot
GeV}\right)^{-0.637}} {\frac{{h}}{500
km}\cdot\frac{{\Phi_{\nu}}}{{\Phi_{\nu}}_o}}
\end{equation}
 This event rate is comparable to UPTAUS one (for comparable fluence) and it may be an
 additional source of Terrestrial Gamma Flashes already observed by GRO
 in last decade \cite{Fargion 2000-2002}. These event rate are
 considered as the detector thresholds for UPTAUS and HORTAUS and
 they are summirized in Table 1 and in Fig. 18 with other present and future
 experimental thresholds.
\section{Appendix A}
 As soon as the altitude $h_1$ and the corresponding energy
$E_{\tau_{h_1}}$ increases the corresponding  air density
decreases. At a too high quota there is no more $X$ slant depth
for any Air-Showering to develop. Indeed its value is :
\begin{displaymath}
X=\int_{\frac{d_u}{2}+c\tau\gamma_t}^{d_1+\frac{d_u}{2}}{{n_0}e^{-\frac{R_{\oplus}}{h_0}{\left[{\sqrt{\left(1-\frac{h_2}{R_{\oplus}}\right)^2+\left({\frac{x}{R_{\oplus}}}\right)^2}}-1\right]}}{dx}}
\end{displaymath}
\begin{equation}
\simeq\int_{\frac{d_u}{2}+c\tau\gamma_t}^{d_1+\frac{d_u}{2}}{n_0e^{-\frac{x^2}{2R_{\oplus}h_0}}dx}\leq{n_0h_0}
\end{equation}
\end{enumerate}
In order to find this critical height $h_{1}$ where the maximal
energy HORTAU terminates  we remind our recent approximation. The
transcendental equation that defines the Tau distance $c\tau$
 has been more simplified in:
\begin{equation}\label{13}
  \int_{0}^{+ \infty} n_0 e^{-\frac{\sqrt{(c\tau+x)^2+R_\oplus^2} - R_\oplus}{h_0}}
   dx \cong n_0 h_0 A
\end{equation}
\begin{equation}\label{14}
  \int_{0}^{+ \infty} n_0 e^{-\frac{(c\tau+x)^2} {2h_0R_\oplus}}
   dx \cong n_0 h_0 A
\end{equation}
\begin{equation}\label{15}
 c\tau = \sqrt{2R_\oplus h_0}
 \sqrt{ln \ffrac{R_\oplus}{c\tau} - ln A }
\end{equation}
Here $A=A_{Had.}$ or $A=A_{\gamma}$ are slow logarithmic
functions of values near unity; applying known empirical laws to
estimate this logarithmic growth (as a function of  the X slant
depth) we  derived respectively for hadronic and gamma UHECR
showers \cite{Fargion 2000-2002}, \cite{Fargion2001a}:
\begin{equation}\label{5}
 A_{Had.}=0.792 \left[1+0.02523 \ln\ffrac{E}{10^{19}eV}\right]
\end{equation}
\begin{equation}\label{5}
 A_{\gamma}=\left[1+0.04343\ln\ffrac{E}{10^{19}eV}\right]
\end{equation}
The solution of the above transcendental equation leads to a
characteristic maximal UHE $c\tau_{\tau}$ = $546 \;km$ flight
distance, corresponding to $E \leq 1.1\cdot 10^{19}eV$ energy
whose decay occurs at height $H_o= 23$ km; nearly 600 Km far from
the horizon it was originated from there on the HORTAUS begins to
shower. At higher quotas the absence of sufficient air density
lead to a suppressed development or to a poor particle shower,
hard to be detected. At much lower quota the same air opacity
filter most of the electromagnetic shower allowing only to muon
bundles and Cherenkov lights to survive at low a somehow $(\leq
10^{-3})$ level.
\section{Appendix B: The UPTAUS area}
The Upward Tau Air-Showers, mostly at PeV energies, might travel a
minimal air depth before reaching the observer in order to
amplify its signal. The UPTAUS Disk Area $A_U$ underneath an
observer at height $h_1$ within a opening angle $\tilde{\theta}_2$
from the Earth Center is:
\begin{equation}
 A_{U}= 2\pi{R_{\oplus}}^2(1 - \cos{\tilde{\theta}_{2}})
\end{equation}
Where the $\sin{\tilde{\theta}_{2}= (x_2/{R_{\oplus}})}$ and
$x_2$ behaves like $x_1$ defined above for HORTAUs. In general the
UPTAUs area are constrained in a narrow Ring (because the mountain
presence itself or because the too near observer distances from
Earth are encountering a too short air slant depth for showering
or a too far and opaque atmosphere for the horizontal
UPTAUs)\cite{Fargion 2002b},\cite{Fargion 2002c}:
\begin{equation}
 A_{U}= 2\pi{R_{\oplus}}^2( \cos{\tilde{\theta}_{3}-\cos{\tilde{\theta}_{2}}})
 \end{equation}
An useful Euclidean approximation is:
\begin{equation}
 A_{U}= \pi {h_1}^2 ({\cot{\theta}_{2}}^2-{\cot{\theta}_{3}}^2)
 \end{equation}
 Where ${\theta}_{2}$, ${\theta}_{3}$ are the outgoing $\tau$
 angles on the Earth surface \cite{Fargion 2000-2002}.

 For  UPTAUs (around $3\cdot10^{15} eV$ energies) these
volumes have been estimated in \cite{Fargion 2000-2002}, assuming
an arrival values  angle$\simeq 45^o- 60^o$ below the horizons.
For two characteristic densities one finds respectively:
$${\Delta M_{eff.}(h_{1}=500 km)(\rho_{Water})}= 5.9~ km^3;$$
$${\Delta M_{eff.}.  (h_{1}=500 km)(\rho_{Rock})}=15.6~ km^3 $$
Their detection efficiency is displayed in last Figure (Fig. 18),
and it exceed by more than an order of magnitude, the future
ICECUBE threshold.

\begin{figure}\centering\includegraphics[width=16cm]{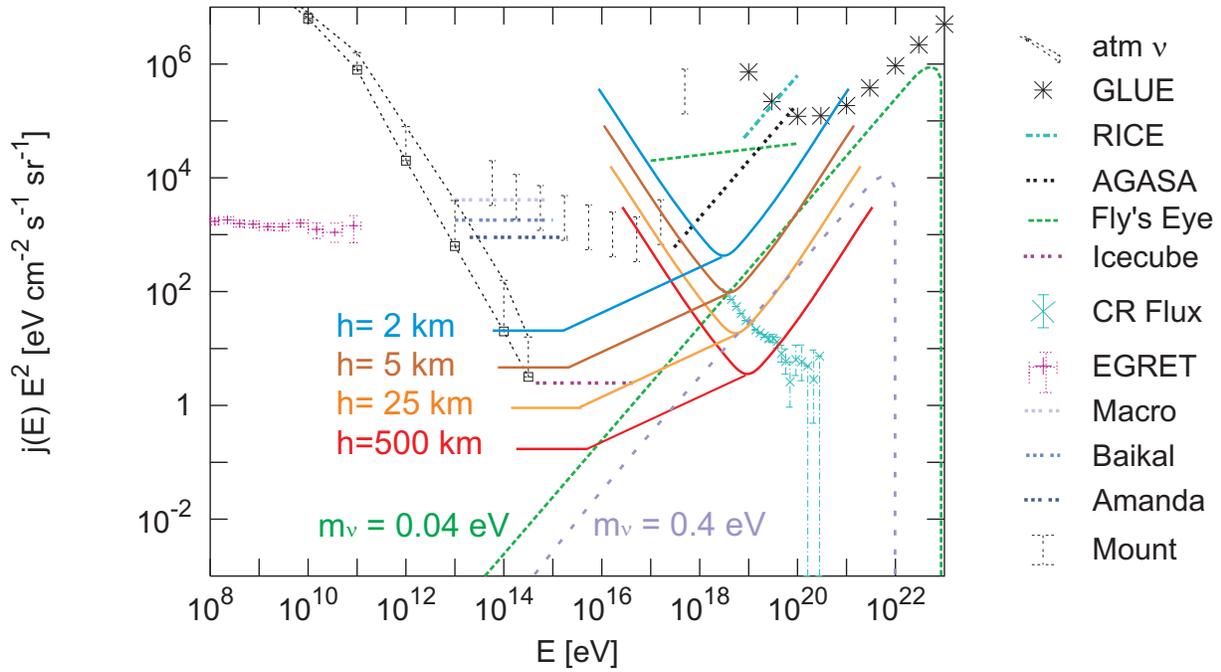}
\caption {UPTAUS (lower bound on the center) and HORTAUS (right
parabolic  curves)  sensibility at different observer heights h
($2,5,25,500 km $) assuming a $km^3$ scale volume (see Table
above)  adapted over a present neutrino flux estimate in Z-Shower
model scenario \cite{Kalashev:2002kx}, \cite{Fargion Mele Salis
1999} for light ($0.4-0.04$ eV) neutrino masses $m_{\nu}$; two
corresponding density contrast has been assumed \cite{Fargion et
all. 2001b}; the lower parabolic bound thresholds are at different
operation height, in Horizontal (Crown) Detector facing toward
most distant horizons edge; these limits are fine tuned (as
discussed in the text); we are assuming a duration of data
records of a decade comparable to the BATSE record data (a
decade). The paraboloid bounds on the EeV energy range in the
right sides  are nearly un-screened by the Earth opacity while
the corresponding UPTAUS bounds  in the  center below suffer both
of Earth opacity as well as of a consequent shorter Tau
interaction lenght in Earth Crust, that has been taken into
account.
\cite{Fargion 2000-2002}, \cite{Fargion 2002b}, \cite{Fargion 2002c}.}
\label{fig:fig18}
\end{figure}
\vspace{-1 cm}
\vspace{1.cm}
\section{Acknowledgements}
The author wishes to thank
P.G. De Sanctis Lucentini, C.Leto, M.De Santis
 for  numerical and technical support
and Prof. G. Salvini and Prof. B. Mele for useful
 discussions and comments.
 \\

\end{document}